\documentclass[11pt]{article}

\usepackage{amsmath}
\usepackage{amsfonts}
\usepackage[paper=a4paper,text={16cm,24cm},centering]{geometry}
\usepackage{graphicx}
\usepackage{ulem}
\usepackage{rotating}

\def\be{\begin{equation}}
\def\ee{\end{equation}}

\begin{document}

\title{Rapidity gap distribution in diffractive deep-inelastic
scattering and parton genealogy}
\author{A.H. Mueller${}^{(1)}$, S. Munier${}^{(2)}$\\
  \footnotesize\it  (1) Department of Physics, Columbia University,
  New York, NY 10027, USA\\
  \footnotesize\it  (2) CPHT, Ecole Polytechnique,
  CNRS, Universit\'e Paris-Saclay, Route de Saclay, 91128 Palaiseau, France\\
}
\date{May 7, 2018}

\maketitle

\begin{abstract}
  We propose a partonic picture for high-mass diffractive dissociation
  events in onium-nucleus scattering,
  which leads to simple and robust predictions for the
  distribution of the sizes of gaps in diffractive dissociation
  of virtual photons
  off nuclei at very high energies.
  We show that the obtained probability distribution
  can formally be identified to the distribution of the decay time of the most
  recent common ancestor
  of a set of objects generated near the edge of a branching
  random walk, and explain
  the physical origin of this appealing correspondence.
  We then use the fact that the diffractive cross section conditioned to a
  minimum rapidity gap size obeys a set of Balitsky-Kovchegov equations
  in order to test numerically our analytical predictions.
  Furthermore, we show how simulations in the framework of a Monte Carlo
  implementation of the QCD evolution support our picture.
\end{abstract}

\section{Introduction}

In scattering processes at energies much larger
than the typical mass of the hadrons (for
a review on high-energy scattering, see Ref.~\cite{Kovchegov:2012mbw}),
a nucleus appears as a weakly
bound system of nucleons, themselves made of dense sets of partons.
Therefore, a classical intuition may lead one to think that in
such processes, the nucleus
would be ripped apart and decay into many hadrons
with probability one.

Instead, it turns out that
the scattering of a hadronic projectile or of a virtual photon
off a target (which may be a nucleus or a proton) leaves the latter intact with a significant
probability, which even tends to $\frac12$ at asymptotically high energies.
This phenomenon is predicted by
basic quantum mechanics: It is just ordinary diffraction. Experimentally,
it was clearly observed
in proton and antiproton collisions
(for reviews, see~\cite{Alberi:1981af,Goulianos:1982vk}),
also in proton-nucleus collisions (the nucleus being
left intact, which is the case we are interested in here)
at CERN~\cite{Bemporad:1971sy}, and later
in virtual photon-proton scattering at DESY HERA~\cite{Ahmed:1995ns,Derrick:1995wv}
(for a review, see Ref.~\cite{Schoeffel:2009aa}).

A priori, diffractive dissociation of a virtual photon
is not very naturally formulated in a traditional perturbative QCD framework,
at variance with the total deep-inelastic scattering cross sections for example.
A continuous effort has been made in trying to set up
a partonic interpretation of diffraction. It was recognized early that
describing diffraction as bare color dipole Fock states of the virtual
photon exchanging a set of globally color-neutral
gluons with the target was a fine theoretical
picture as a starting point~\cite{Nikolaev:1991et},
which can explain qualitatively the relatively large
fraction of diffractive events (about 10\%) observed in the HERA
data~\cite{Buchmuller:1995qa}.
Different authors have refined this picture
by incorporating small-$x$ QCD evolution or/and shadowing corrections
(see e.g.~\cite{Mueller:1994jq,Bialas:1995bs,Gotsman:1999vt}),
others instead have developed
a QCD factorization scheme for diffractive processes, introducing
diffractive parton densities
(see e.g.~\cite{Collins:1997sr}).

A major phenomenological
success of saturation models such as the Golec-Biernat
and W\"usthoff model is that they have been able to describe both
diffractive and inclusive cross sections in the very same
formalism~\cite{GolecBiernat:1998js}.
This success fostered many other developments, phenomenological
(see e.g. Ref.~\cite{GayDucati:2001zf})
as well as theoretical.
Among the latter, of particular interest for us
are the works aimed at establishing and studying
nonlinear evolution equations for high-mass diffractive
cross sections~\cite{Kovchegov:1999ji,Levin:2001yv,
  Levin:2001pr,Levin:2002fj,Kovner:2001vi,Hentschinski:2005er,Hatta:2006hs}.

In the present paper, we derive robust new theoretical predictions for
the distribution of rapidity gaps in the diffractive dissociation of a
small-size onium (which in an experiment may be sourced
by virtual photons picked from the field
of a lepton) off a large nucleus.
We show that this distribution is the same
-- up to the overall normalization -- as the
distribution of the age of the most recent common ancestor of extreme
particles in branching random walks. 
We also rederive evolution equations for high-mass diffractive cross sections:
We recover the fact that they
may be obtained as the solution of Balitsky-Kovchegov equations.
A numerical implementation of the latter
enables us to check our analytical
calculation of the distribution of the size of the rapidity gap.

Our paper is organized as follows.
In Sec.~\ref{sec:picture}, after a short review of
high-energy evolution and scattering, we introduce our picture
of diffraction, and explain how it connects to
the general problem of ancestry in branching random walks.
In Sec.~\ref{sec:BK}, we show how to formulate rigorously diffractive cross sections as solutions
to a system of Balitsky-Kovchegov equations,
which enables accurate numerical checks of our expression for the
distribution of gaps. Our numerical results are presented in Sec.~\ref{sec:numerics}.


\section{\label{sec:picture}Gap distribution and statistics of common ancestors}

\subsection{Short review on high-energy evolution and scattering}

In this section and throughout this paper,
we will consider an onium of transverse size $x_{01}$ interacting with a large nucleus.

\subsubsection{Diagonal $S$-matrix elements and their evolution}

Most generally, the interaction
cross section can be decomposed in its elastic and inelastic components
$\sigma_\text{el}$ and $\sigma_\text{in}$, the total cross section
$\sigma_\text{tot}$ being the sum of the latter two.
(By ``cross section'' we always mean the dimensionless quantity ``cross section
per unit surface in transverse space''
parametrized by the impact parameter~$b$,
which for simplicity, we denote by $\sigma$ instead of
$d\sigma/d^2b$.)
These cross sections may be expressed
with the help of the $S$-matrix element $S(x_{01})$
for the elastic scattering
of the dipole off the nucleus:
\be
{\sigma_\text{el}}(x_{01})=\left|1-S(x_{01})\right|^2
\ ,\quad
{\sigma_\text{in}}(x_{01})=1-\left|S(x_{01})\right|^2
\ ,\quad
  {\sigma_\text{tot}}(x_{01})=2\left[1-\text{Re}\,S(x_{01})\right].
  \label{eq:sigma_S}
\ee
At very high energies, $S$ is essentially real and ranges between
$0$ and $1$, so the modulus and real part can be dropped
in the above formulae.
It is also useful to introduce the scattering amplitude $T(x_{01})=1-S(x_{01})$.

These formulae are general, not dependent on the
microscopic theory.
In QCD, an equation for the change in $S$ as the center-of-mass
energy (or, equivalently, the total rapidity~$y$)
is increased can be written down. It
is the Balitsky-Kovchegov (BK) equation~\cite{Balitsky:1996ub,Kovchegov:1999yj},
established in the framework of the color dipole model~\cite{Mueller:1993rr}
and which reads
\be
\frac{\partial S(x_{01},y)}{\partial y}=\bar\alpha \int
\frac{d^2x_2}{2\pi}\frac{x_{01}^2}{x_{02}^2 x_{12}^2}
\left[S(x_{02},y)S(x_{12},y)-S(x_{01},y)\right],
\label{eq:BK_S}
\ee
where we made the $y$-dependence explicit, and we introduced the usual notation
$\bar\alpha=\alpha_s N_c/\pi$.
The initial condition is a function of $x_{01}$
representing the $S$-matrix element at some starting rapidity,
say $y=0$. For the latter, one can use the McLerran-Venugopalan (MV)
model~\cite{McLerran:1993ni}, representing the interaction of a dipole of size $x_{01}$ with
the nucleus characterized by the momentum scale $Q_\text{MV}$:
\be
S_\text{MV}(x_{01})=\exp\left[-\frac{x_{01}^2 Q_\text{MV}^2}{4}
  \ln\left(e+\frac{4}{x_{01}^2\Lambda^2}
  \right)
  \right].
\label{eq:MV_S}
\ee
$\Lambda$ is the QCD scale.
We see that $S_\text{MV}$ is then a function smoothly
and monotonously connecting
$S_\text{MV}=1$ for $x_{01}\rightarrow 0$ to $S_\text{MV}=0$ for $x_{01}\rightarrow +\infty$,
with a sharp transition occurring around the size $2/Q_\text{MV}$.

The solution to the BK equation is not known, but some of its essential properties
have been derived, see Ref.~\cite{Mueller:2002zm,Munier:2003vc,Munier:2003sj}.
In particular, asymptotically for large $y$, it converges
to a traveling wave, which at fixed $y$ is again a smooth function of $x_{01}$
connecting~1 and~0, whose evolution with $y$ amounts to a mere translation in $x_{01}$.
The transition between $S=1$ and $S=0$
occurs at some $y$-dependent size $2/Q_s(y)$, where $Q_s(y)$
is the saturation momentum, defined for example by requiring that
\be
S(x_{01}=2/Q_s(y),y)=\frac12.
\ee
With an initial condition such as the MV model,
the analytic expression of the amplitude $T=1-S$ around the transition region
reads
\be
T(x_{01},y)=c_T \ln\frac{1}{x_{01}^2Q_s^2(y)}
\left[x_{01}^2Q_s^2(y)\right]^{\gamma_0}
\exp\left\{
  -\frac{\ln^2 [x_{01}^2Q_s^2(y)]}{2\bar\alpha y\chi''(\gamma_0)}
  \right\}.
  \label{eq:Tscaling}
\ee
(A precise definition of the validity domain will be given in Eq.~(\ref{eq:scaling_region}) below).
$c_T$ is a constant and
the saturation momentum reads, for large $y$,
\be
Q_s^2(y)=Q_{\text{MV}}^2
\frac{e^{\bar\alpha y\chi'(\gamma_0)}}
     {({\bar\alpha y})^{3/(2\gamma_0)}},
\ee
up to an overall constant of order one not explicitely written here,
which depends on the very definition of the saturation scale.
The function
\be
\chi(\gamma)=2\psi(1)-\psi(\gamma)-\psi(1-\gamma)
\ee
is up to a factor $\bar\alpha$
the usual eigenvalue of the BFKL kernel corresponding to the
eigenfunction $x_{01}^{2\gamma}$.
 $\gamma_0$ is defined to be the solution to the equation
$\chi(\gamma_0)/\gamma_0=\chi'(\gamma_0)$.
In numbers,
\be
\gamma_0=0.627549\cdots,\quad
\chi(\gamma_0)=3.0645\cdots,\quad
\chi''(\gamma_0)=48.5176\cdots
\ee

The forms of $T$ and of the saturation momentum
do not depend on
the details of the initial condition, up, of course,
to overall constants and subasymptotic terms
in the rapidity. The expression~(\ref{eq:Tscaling})
is valid in the
so-called (geometric) scaling region \cite{Stasto:2000er}
but nevertheless for $x_{01}\ll 1/Q_s(y)$.
The scaling region is defined
as the range in $x_{01}$ in which $T$,
which is a priori a function of the two variables $x_{01}$
and $y$, becomes effectively a function of the
single composite variable $x_{01}Q_s(y)$ only.
Hence parametrically, $x_{01}$
needs to obey the following inequalities:
\be
1<|\ln [x_{01}^2 Q_s^2(y)]|\leq\sqrt{\bar\alpha y\chi''(\gamma_0)}.
\label{eq:scaling_region}
\ee
In this region, the scattering is weak ($T\ll 1$)
but the value of the amplitude is influenced by the presence
of a saturated nucleus, which, technically, acts as an absorptive
boundary.

\subsubsection{Interpretation of the BK equation}

Let us interpret physically the BK equation.

Assume that the total rapidity available for the
scattering is $Y$. Integrating Eq.~(\ref{eq:BK_S}) up
to $y=Y$, we get $T(x_{01},Y)$ which we may view as the elastic
scattering amplitude for the interaction of an {\it elementary} dipole
of size $x_{01}$ off a nucleus evolved to rapidity $Y$. This
is the dipole restframe picture.
In this frame, the nucleus looks the same in each event: It is a dense
set of gluons, whose density grows {\it deterministically} with the rapidity. The
density of the gluons in the nucleus
determines the probability amplitude for the onium to interact with it.

One can have a completely different interpretation
by going for example to the restframe of the nucleus. (See Ref.~\cite{Mueller:2014fba}
where these ideas were developed).
In that frame, the onium carries
the full rapidity $Y$ and thus is in a highly-evolved quantum
state at the time of its interaction with the nucleus.
This quantum state is conveniently represented by
a set of dipoles, differing from an event to another one,
constructed stochastically through
a $1\rightarrow 2$ branching diffusion
process whose characteristics
are encoded in the kernel of the BK equation.
Indeed, the probability that a dipole of size~$x_{01}$ splits by emission of a
gluon at transverse position $x_2$ up to $d^2x_2$ into two dipoles
of sizes $x_{02}$ and $x_{12}$ when one increases its rapidity
by the infinitesimal quantity $d\tilde y$ reads~\cite{Mueller:1993rr}
\be
\text{proba}(x_{01}\rightarrow x_{02},x_{12})=
\bar\alpha \,d\tilde y\, \frac{d^2 x_2}{2\pi}\frac{x_{01}^2}{x_{02}^2 x_{12}^2}.
\label{eq:dipole_splitting}
\ee
The scattering amplitude $T(x_{01},Y)$ is then
related to the probability $P(x_{01},Y|1/Q_\text{MV})$ to find {\it at least} one dipole
in the onium Fock state 
at rapidity $\tilde y=Y$
whose size is larger than $R=1/Q_\text{MV}$.
The BK equation then appears as an equation for the statistics of
the size of the largest dipole in a branching diffusion process.

More precisely, $P$
solves the BK equation written in the form
\begin{multline}
\frac{\partial P(x_{01},\tilde y|R)}{\partial \tilde y}=\bar\alpha \int
\frac{d^2x_2}{2\pi}\frac{x_{01}^2}{x_{02}^2 x_{12}^2}
[P(x_{02},\tilde y|R)+P(x_{12},\tilde y|R)-P(x_{01},\tilde y|R)\\
  -P(x_{02},\tilde y|R)P(x_{12},\tilde y|R)].
\label{eq:BK_T}
\end{multline}
This equation is the same equation as the one solved by $T=1-S$.
The initial conditions however differ: The one for $T$
is e.g. the MV model, while the one for $P$ is a mere
Heaviside distribution
\be
P(x_{01},\tilde y=0|R)=\theta\left(
\ln\frac{x_{01}^2}{R^2}
\right)
\ee
when the initial state of the onium consists in a single dipole of size $x_{01}$.
Although the initial conditions are different, the asymptotic
large-rapidity solutions fall into
the very same universality class.
It is this mathematical property which enables the identification
of the asymptotics of the
scattering amplitude of a dipole of size $x_{01}$ off a nucleus characterized
by the saturation momentum $Q_\text{MV}$ at total rapidity $Y$ with
the probability of finding (at least)
a dipole of size larger than $1/Q_\text{MV}$ at rapidity $Y$
in an initial onium of size $x_{01}$:
\be
T(x_{01},Y)\underset{\text{large $Y$}}{\simeq}P(x_{01},Y|1/Q_\text{MV}).
\ee
For completeness and because we will use it below,
let us write explicitely the expression of $P$,
which is tantamount to the expression of $T$ up to the appropriate substitutions:
\be
P(x_{01},\tilde y|R)=c_P \ln\frac{R^2}{x_\perp^2(\tilde y)}
\left[\frac{x_\perp^2(\tilde y)}{R^2}\right]^{\gamma_0}
\exp\left\{
-\frac{\ln^2 [R^2/x_\perp^2(\tilde y)]}{2\bar\alpha \tilde y\chi''(\gamma_0)}
\right\}
\label{eq:Pscaling}
\ee
(compare to Eq.~(\ref{eq:Tscaling})).
$c_P$ is a constant and
\be
x_\perp^2(\tilde y)=x_{01}^2 \frac{e^{\bar\alpha\tilde y\chi'(\gamma_0)}}
{(\bar\alpha\tilde y)^{3/(2\gamma_0)}},
\label{eq:x_perp}
\ee
up to a constant which again is a matter of definition.

The interpretation of $x_\perp(\tilde y)$ is the following.
It is the size for which the probability to have a dipole
of that size in the Fock state of the onium at rapidity~$\tilde y$
is some predefined number, say~1.
If one draws the histogram of the dipole sizes at rapidity~$\tilde y$
in a particular event,
$x_\perp$ is the {\it expected} position of its tip (towards
large sizes).
Consequently, dipoles of sizes much larger than $x_\perp$
can only stem from rare fluctuations in the stochastic evolution.

The probability of having such fluctuations in a
realization of the onium evolution is precisely given
by Eq.~(\ref{eq:Pscaling}). In the same way as
for $T$, there is a region defined by Eq.~(\ref{eq:scaling_region})
(with the substitutions $y\rightarrow\tilde y$,
$Q_s(y)\rightarrow 1/x_\perp(\tilde y)$)
in which the probability to find a fluctuation
obeys a scaling law, and
out of which (for very large $R$) $P$ is strongly
suppressed by the Gaussian factor in the log of the sizes
appearing in Eq.~(\ref{eq:Pscaling}).

\subsection{Rapidity gaps and large fluctuations}

We are now in a position to introduce our picture of diffraction.
We shall start by defining diffractive versus elastic
events, before moving on to the microscopic description of diffractive
events and eventually to the quantitative predictions
which directly follow.

\subsubsection{\label{sec:el_diff}Elastic and diffractive events}

Elastic events are, at least theoretically, straightforward to define:
The particles in the final state are the same as the ones in the initial
state; Only their momenta are redistributed.
In order to observe a significant fraction of elastic events,
one needs that the $S$-matrix be close to 0.
Indeed, in this case, according to Eq.~(\ref{eq:sigma_S})
\be
\sigma_\text{el}\simeq\sigma_\text{in}\simeq
\sigma_\text{tot}/2,
\ee
and the
ratio $\sigma_\text{el}/\sigma_\text{tot}$ is maximum.
These equalities are characteristic of the scattering of quantum particles off a black
disk: The inelastic cross section corresponds to the absorption by the disk,
while the elastic cross section is due to the particles which are diffracted in its shadow.

For a single elementary dipole (i.e. an onium in its ground state) that
scatters off a nucleus,
$S=0$ is verified whenever the size of the dipole is much larger than the
inverse saturation momentum $1/Q_s(Y)$ of the nucleus ($Y$ is
the rapidity of the nucleus in the frame of the dipole, namely the total
available rapidity). In this case, elastic scattering is observed in half of the events.

If instead the size of the onium is very small compared to the inverse saturation momentum,
then $S\sim 1$, and the total cross section is dominated
by the inelastic events:
\be
\sigma_\text{in}=(1+S)(1-S)\simeq 2(1-S)=\sigma_\text{tot}\gg (1-S)^2
=\sigma_{\text{el}}
\ee
Hence one may think that a small onium (small compared to $1/Q_s(Y)$)
would almost exclusively trigger
inelastic events since its inelastic cross section
is small, but its elastic cross section
is even much smaller.
This is however not true, because in a high-energy scattering,
the onium does not interact as a bare dipole
state but through complicated quantum fluctuations,
and the latter may interact
elastically with a significant probability.
Such realizations of the onium state
do not result in elastic events,
but in diffractive ones, which we shall now define more precisely.\\

\begin{figure}
  \begin{center}
  \begin{tabular}{cc}
  \includegraphics[width=0.7\textwidth]{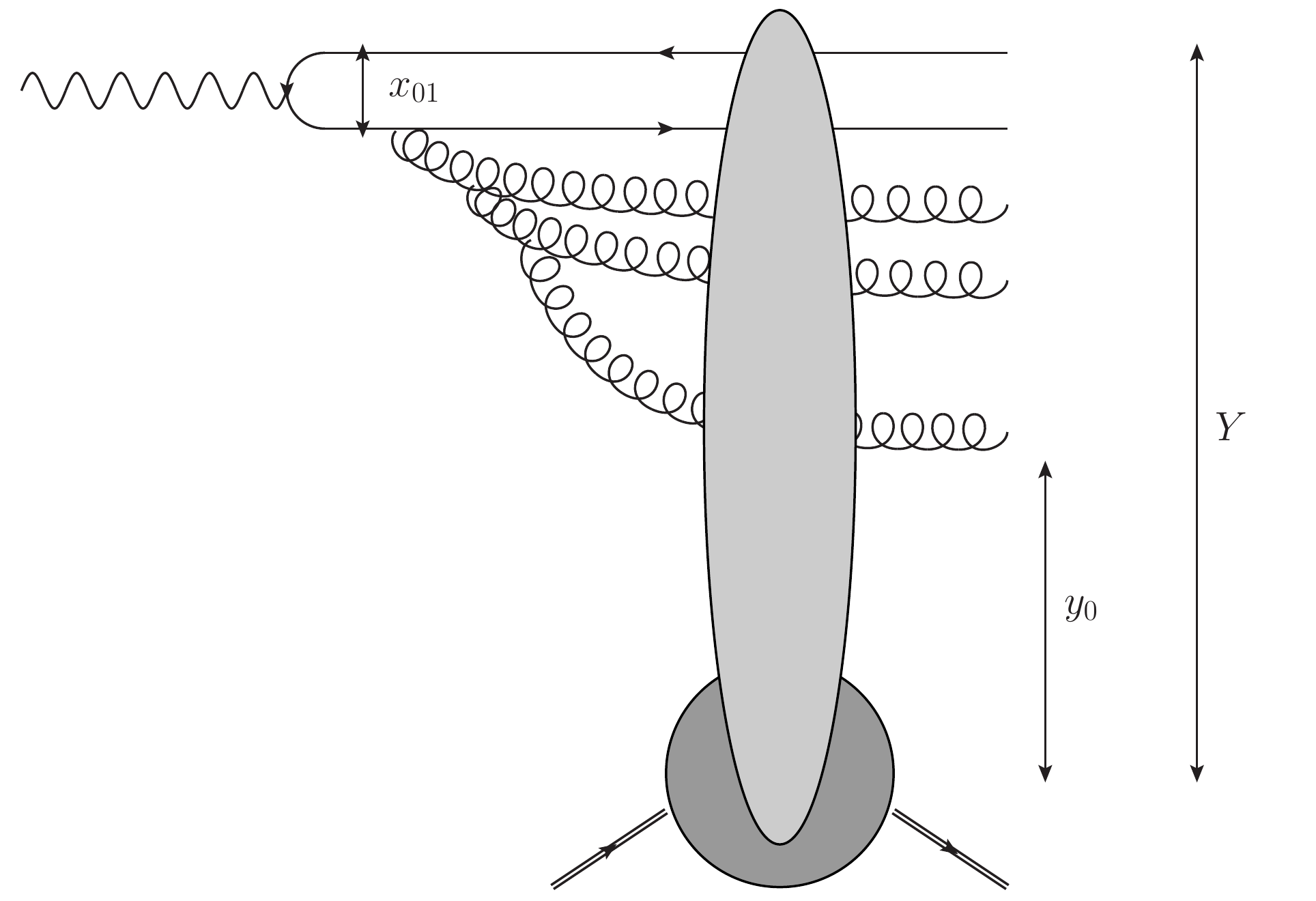}
  \end{tabular}
  \end{center}
  \caption{\small \label{fig:diffraction_diag}
    Graph contributing to the amplitude for high-mass diffraction
    in deep-inelastic scattering, with a rapidity gap $y_0$.
    The virtual photon converts to an onium of size $x_{01}$, which scatters
    through a Fock state made of the initial $q\bar q$ pair together with
    soft gluons.
    The elongated blob represents the exchange of a color singlet   
    between the state of the onium and the nucleus resulting
    in an elastic scattering.
}
\end{figure}

To define diffractive events, we do not require
that both interacting particles remain of the same nature.
We just require that the nucleus does not
completely break up in the scattering. It may remain strictly intact or scatter
to an excited state, which subsequently returns
to an energy minimum through fission.
The scattering 
is coherent or quasi-elastic,
at the level of the nucleus.
In any case,
there is a region in rapidity (namely in angle in a detector) around
the decay products of the nucleus in which no particle is seen.
Requiring that there be a rapidity gap is a good practical definition,
and was a striking signature of diffractive events at HERA.
The onium instead may undergo a transition to a
complicated system of many hadrons
which materialize in the final state (see Fig.~\ref{fig:diffraction_diag}).
This is what happens if we ask for a high-mass diffractive
final state.\footnote{The elastic scattering of the onium
  contributes to low-mass diffraction,
  and has been studied in the context of deep-inelastic
  scattering by many groups,
see e.g. Ref.~\cite{Bartels:1996ne,Bialas:1996tn,GolecBiernat:1998js}.
This is not our focus here.}

Coming back to onium-nucleus scattering and
with a view at sorting out the diffractive dissociative events,
it is useful to picture the scattering
in a frame in which the onium is not at rest: Let us give
it the rapidity $\tilde y_0\gg 1$, while
the nucleus takes the remaining available rapidity $y_0=Y-\tilde y_0$.
Then the onium does not interact
as a bare $q\bar q$ dipole state, but as a quantum state
made of many gluons (conveniently represented
by a set of dipoles). In such quantum states, there may be a few unusually large
dipoles, of sizes greater than $1/Q_s(y_0)$. These dipoles interact with
the nucleus with a probability of order~1. If furthermore their size lie deep
in the saturation region, the interaction with the nucleus
is {\it elastic} in half of the events. This results in a rapidity gap
whose size is of the order of the remaining rapidity, namely $Y-\tilde y_0=y_0$.
At the same time,
because the different components of the Fock state interact
all differently,
the coherence is broken at the level of the onium and the partons
present in its Fock state at rapidity $\tilde y_0$
materialize in
the form of a hadronic system in the fragmentation
region of the latter.
This is precisely diffractive dissociation.

So the key for describing diffractive events is a proper understanding
of the event-by-event
fluctuations of the
large-dipole component of the
onium quantum state.

\subsubsection{Fluctuations in onium evolution}

For $x_{01}$ small enough with respect to $1/Q_s(Y)$,
the mechanism for the production of an unusually large dipole
(that crosses the saturation
boundary at rapidity $\tilde y_0=Y-y_0$ with respect to the onium,
up to typically one unit) in the Fock state of an onium
was studied in detail in Ref.~\cite{Mueller:2014fba,Mueller:2014gpa}.
There, we identified two types of fluctuations that may occur in the course of the
rapidity evolution:
The {\it front fluctuations} and the
{\it tip fluctuations}.

In the beginning of the onium evolution, i.e. at
low rapidities~$\tilde y$,
the state is dilute and thus subject to fluctuations which
strongly determine the size of the largest dipole in the event at any
later rapidity.
We shall call this size ``position of the tip of the front'' by reference
to the traveling wave language,
and the kind of fluctuations which lead to this effect {``front
  fluctuations''}.

At rapidity $\tilde y_0$ (up to $\Delta \sim 1/\bar\alpha$), a
fluctuation at the tip of the front may happen, containing one
or a few dipoles typically larger by an order of magnitude with
respect to the largest dipole in the absence of the fluctuation.
The latter fluctuation
is short-lived because it is rapidly absorbed by the bulk of
the front, which moves with a velocity larger than that
of the tip of the front stemming from the fluctuation.
This is what we call a ``tip fluctuation'', and is quite
different in nature from
the front fluctuations: The memory of the latter
is conserved throughout the evolution, whereas the former
moves stochastically the tip of the front forward
only at the very rapidity at which it occurs.

The most
favorable scenario for creating a few unusually large
dipoles around the rapidity~$\tilde y_0$
is to combine a front fluctuation occurring
at low rapidities $\tilde y\simeq 0$
at which the onium still is a dilute state 
with a tip fluctuation at~$\tilde y_0$.

Actually, it will be an essential ingredient of our calculation that the dipoles 
which are within the saturation region of the nucleus at
rapidity $\tilde y_0$ come from this very combination of front and tip fluctuations.
The analytical arguments supporting this scenario are quite technical:
Therefore, we leave them for the Appendix.

\begin{figure}
  \begin{center}
  \begin{tabular}{cc}
  \includegraphics[width=0.9\textwidth]{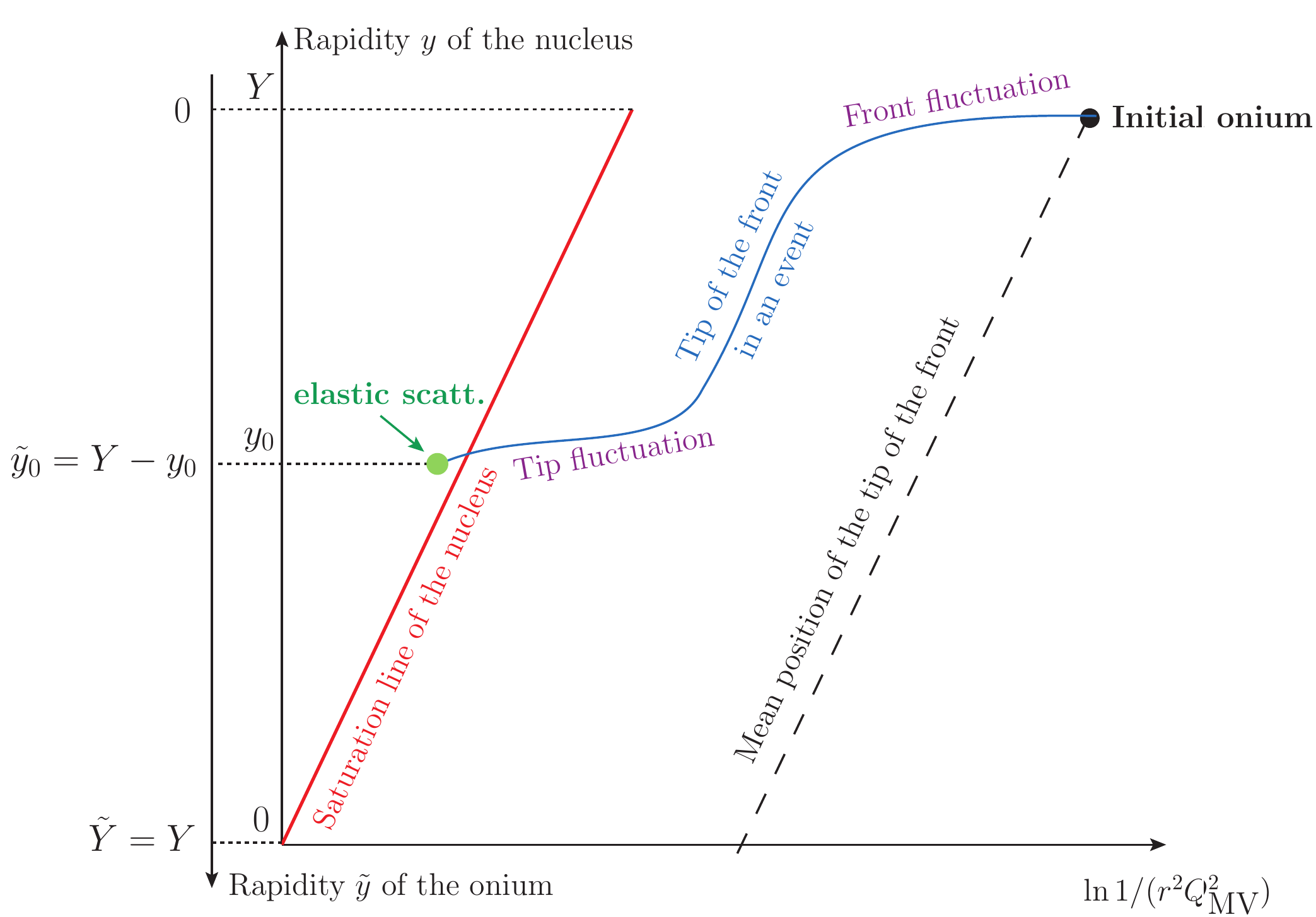}
  \end{tabular}
  \end{center}
  \caption{\small \label{fig:picture}Schematic picture of
    a diffractive event with a rapidity gap of size $y_0$.
    The vertical axis oriented upwards is the rapidity of the nucleus,
    ranging from
    0 to the total available rapidity~$Y$. The rapidity
    of the onium is marked on the downward axis.
    The saturation line of the nucleus of equation $rQ_s(y)\sim 1$
    is shown (continuous straight line),
    as well as the size of the largest dipole in
    the Fock state of the onium in this particular realization of the evolution
    (also called ``position of the tip of the front''; continuous curly line).
    Its shape is determined by a front fluctuation and a tip
    fluctuation, which together generate a few unusually large dipoles
    (compared to the mean position of the tip $r=x_\perp(\tilde y)$; dashed line)
    at rapidity $y_0$.}
\end{figure}

\subsubsection{Quantitative evaluation of the diffractive cross section}

The scenario for diffractive events is illustrated schematically in Fig.~\ref{fig:picture}.
Let us translate this picture into an actual quantitative
evaluation of the diffractive cross section.

In our picture, a given event has a rapidity gap of size $y_0$
when the Fock state of the onium in this
event contains at least one
dipole of size larger than $1/Q_s(y_0)$ at rapidity $\tilde y_0=Y-y_0$,
namely if there is a fluctuation at that rapidity which generates a large dipole.

This means that $d\sigma_\text{diff}/dy_0$ is identical to the probability
$P(x_{01},\tilde y_0|1/Q_s(y_0))$ introduced above Eq.~(\ref{eq:BK_T}),
\be
\frac{d\sigma_\text{diff}}{dy_0}
=P(x_{01},\tilde y_0|1/Q_s(y_0)).
\label{eq:sigma_diff=P}
\ee
As explained before, $P$
solves the BK equation. Therefore, the following expression
holds for the diffractive
cross section in the scaling region
and at its border (see Eq.~(\ref{eq:Pscaling})):
\be
\frac{d\sigma_\text{diff}}{dy_0}=c_\text{diff}
\left[{x_\perp^2(\tilde y_0)Q_s^2(y_0)}\right]^{\gamma_0}
\ln\frac{1}{x_\perp^2(\tilde y_0)Q_s^2(y_0)}
\exp\left\{
  -\frac{\ln^2 [x_\perp^2(\tilde y_0)Q_s^2(y_0)]}
  {2\bar\alpha \tilde y_0\chi''(\gamma_0)}
  \right\},
  \label{eq:dsigma1}
\ee
where $c_\text{diff}$ is a constant.
This formula applies in the scaling region, away from the
deep saturation region,
which puts the following restrictions on the
values of the rapidity and the size of the initial
onium (see Eq.~(\ref{eq:scaling_region})):
\be
1<|\ln [x_{\perp}^2(\tilde y_0)Q_s^2(y_0)]|
\leq\sqrt{\bar\alpha \tilde y_0\chi''(\gamma_0)}.
\label{eq:condition_scaling}
\ee
In order to isolate the $y_0$ dependence,
we can rewrite Eq.~(\ref{eq:dsigma1}) as
\be
\boxed{
\frac{d\sigma_\text{diff}}{dy_0}
=c_\text{diff} \left[\frac{\bar\alpha Y}{\bar\alpha y_0 \bar\alpha(Y-y_0)}
  \right]^{3/2}\exp\left\{-\frac{\ln^2[x_{01}^2Q_s^2(Y)]}
         {2\bar\alpha(Y-y_0)\chi''(\gamma_0)}\right\}
\times  \left[x_{01}^2Q_s^2(Y)
  \right]^{\gamma_0}\ln\frac{1}{x_{01}^2Q_s^2(Y)}},
\label{eq:diff}
\ee
where we have furthermore assumed\footnote{%
  Actually, it would be enough not to approach
  the endpoints
  $y_0=0$ and $y_0=Y$ by less than $1/\bar\alpha$, in such
  a way that the r.h.s. of Eq.~(\ref{eq:approx_log})
  can be considered slowly varying
  compared to the l.h.s.
  }
\be
\left|\ln[x_{01}^2Q_s^2(Y)]\right|\gg
\left|\ln \frac{\bar\alpha Y}
          {\bar\alpha y_0\,\bar\alpha(Y-y_0)}
          \right|
\label{eq:approx_log}
\ee
in order to be able to substitute
$\ln [x_\perp^2(Y-y_0)Q_s^2(y_0)]$ with $\ln [x_{01}^2Q_s^2(Y)]$.
This condition in turn requires that $y_0$ cannot be too close neither to $0$
nor to $Y$.
Note that within this approximation,
the condition~(\ref{eq:condition_scaling})
translates into a condition on the size of the initial onium:
\be
1<|\ln [x_{01}^2Q_s^2(Y)]|\leq\sqrt{\bar\alpha Y\chi''(\gamma_0)}.
\ee
These inequalities mean that we need to pick the initial dipole away
from the saturation region (this is what the
leftmost inequality means) but
in the scaling region or at its border (rightmost inequality), in such
a way that the rapidity evolution of the dipole is always driven by
the eigenvalue of the BFKL kernel $\bar\alpha\chi(\gamma\simeq\gamma_0)$.

Equation~(\ref{eq:diff}) is the main result of our paper.
For fixed~$Y$ and~$x_{01}$, it gives the distribution of the rapidity
gap in diffractive events, up to a constant
which we were not able to determine analytically.

It is also interesting to compare the diffractive
and total cross sections. To this aim,
let us go deep into the scaling region, which
allows us to neglect the Gaussian suppression
factor in Eq.~(\ref{eq:diff}). Recalling that
\be
\sigma_\text{tot}=c_\text{tot}
  \left[x_{01}^2Q_s^2(Y)
    \right]^{\gamma_0}\ln\frac{1}{x_{01}^2Q_s^2(Y)},
\ee
we arrive at
\be
\boxed{
  \frac{1}{\sigma_\text{tot}}
  \frac{d\sigma_\text{diff}}{dy_0}
  =\frac{c_\text{diff}}{c_\text{tot}}
  \left[\frac{\bar\alpha Y}{\bar\alpha y_0 \bar\alpha(Y-y_0)}
    \right]^{3/2}}.
\label{eq:main_result}
\ee

The $y_0$-dependence is completely determined by this formula.
However, we do not see how we may determine
the overall constant $c_\text{diff}/c_\text{tot}$ from our approach.

We are now going to explain how this is related to genealogies
in branching random walks.


\subsection{Parton genealogy}

In this section, we draw a parallel between diffraction in hadronic
physics and ancestry
in general branching-diffusion processes.

Let us go to the restframe of the nucleus, in which the whole
evolution is in the onium. In a given event, a few
dipoles in the Fock state of the onium
eventually interact with the nucleus, typically the ones
that have a size larger than the inverse saturation momentum $Q_\text{MV}^{-1}$
of the nucleus.

We chose the initial onium in such a way that the mean position of the tip
of the front was far enough from the saturation boundary of the nucleus that
diffractive scattering was due to an unusually large fluctuation in the
course of the evolution.
It is this fluctuation which generates, through
further evolution, the dipoles that
eventually scatter off the nucleus in the nucleus restframe.
Hence with high probability,
this particular fluctuation contains the most recent common ancestor
of the dipoles that interact.

A similar but not identical problem has recently been addressed by
Derrida and Mottishaw in the context of the so-called Generalized Random
Energy Model (GREM)~\cite{DM}, from
which they got results that they assumed applied to branching random walks.

The kind of problem addressed there was the following. Consider e.g.
a one-dimensional branching random walk.
Think of branching Brownian motion for example: Independent particles
diffuse on a line parametrized by the real variable~$x$ in time~$t$, and each of them
may independently generate two offspring
according to a Poisson process in time of constant intensity.
Starting with one particle at~$t=0$,
at the final time $t=T$, each realization consists in a
given finite number of particles, extending over a finite region in $x$.
Let us pick the $n$ leftmost particles in each event~($n=2,3,\cdots$)
and ask the following question: What is the
time $t_0$ at which their most recent common ancestor split?

The answer was found (actually conjectured by extrapolating a calculation
in the GREM)
in Ref.~\cite{DM}. The probability density of $t_0$ reads
\be
p(t_0)=\frac{1}{\gamma_0}
\sqrt{\frac{1}{2\pi\chi''(\gamma_0)}}
\left(\frac{T}{t_0(T-t_0)}\right)^{3/2}
\ee
Note that the overall normalization factor could also be
computed in that
context. $\chi(\gamma)$ is the eigenvalue function of the
kernel of the
equivalent of the linearized BK equation and $\gamma_0$ 
is defined to be the solution of the equation $\chi(\gamma_0)/\gamma_0=\chi'(\gamma_0)$.
Up to the replacements of the evolution variables
$\bar\alpha Y\leftrightarrow T$,
$\bar\alpha y_0\leftrightarrow t_0$ and up to the overall
normalization,
this is exactly the same answer as the one
we obtained for the gap size dependence of the diffractive cross section,
see Eq.~(\ref{eq:main_result}).

We shall test this analogy numerically below in Sec.~\ref{sec:stochastic_formulation}.
Before, we are going to introduce the Good-Walker formulation which is very useful
to set up a numerical calculation of the diffractive cross section.


\section{\label{sec:BK}Diffraction from the BK equation}

In this section, we give a concise derivation of
an equation for the diffractive cross section which was first established
in Ref.~\cite{Kovchegov:1999ji} (see also Ref.~\cite{Kovchegov:2012mbw}, Chap.~7.2),
and further studied in Ref.~\cite{Levin:2001yv,Levin:2001pr}.
The authors of Ref.~\cite{Hatta:2006hs} have also reestablished this equation,
in a way which is very close to ours.

\subsection{Good-Walker formulation of diffraction}

Good and Walker gave an elegant general framework for diffractive dissociation
\cite{Good:1960ba}, which will enable us to write exact equations for diffraction
in the framework of the dipole model.
Its essential ingredient is the expansion of
the state of a projectile in terms of the eigenstates
of the interaction of a given target off which it scatters.

\begin{figure}
  \begin{center}
  \begin{tabular}{cc}
  \includegraphics[width=0.36\textwidth]{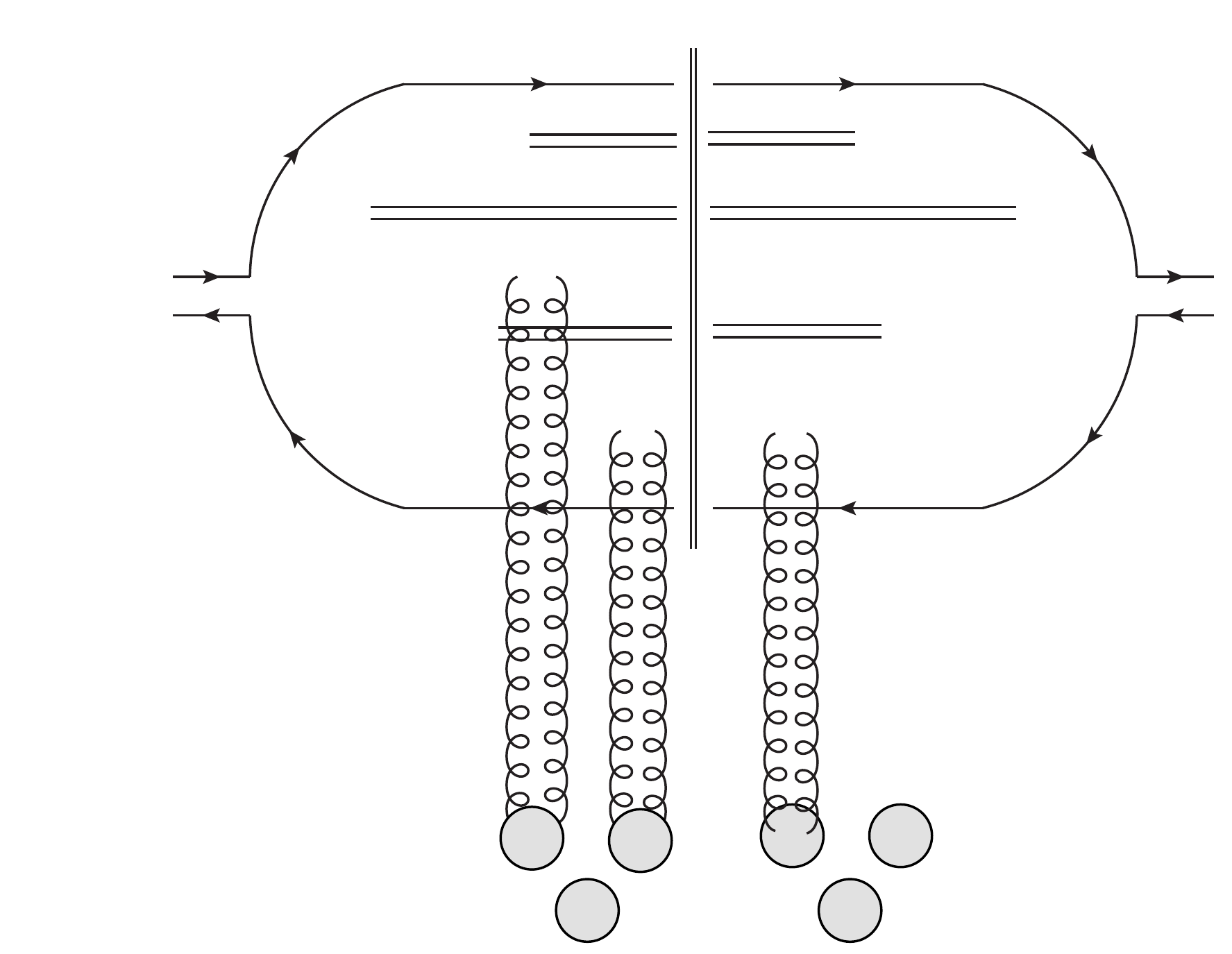} &
  \includegraphics[width=0.54\textwidth]{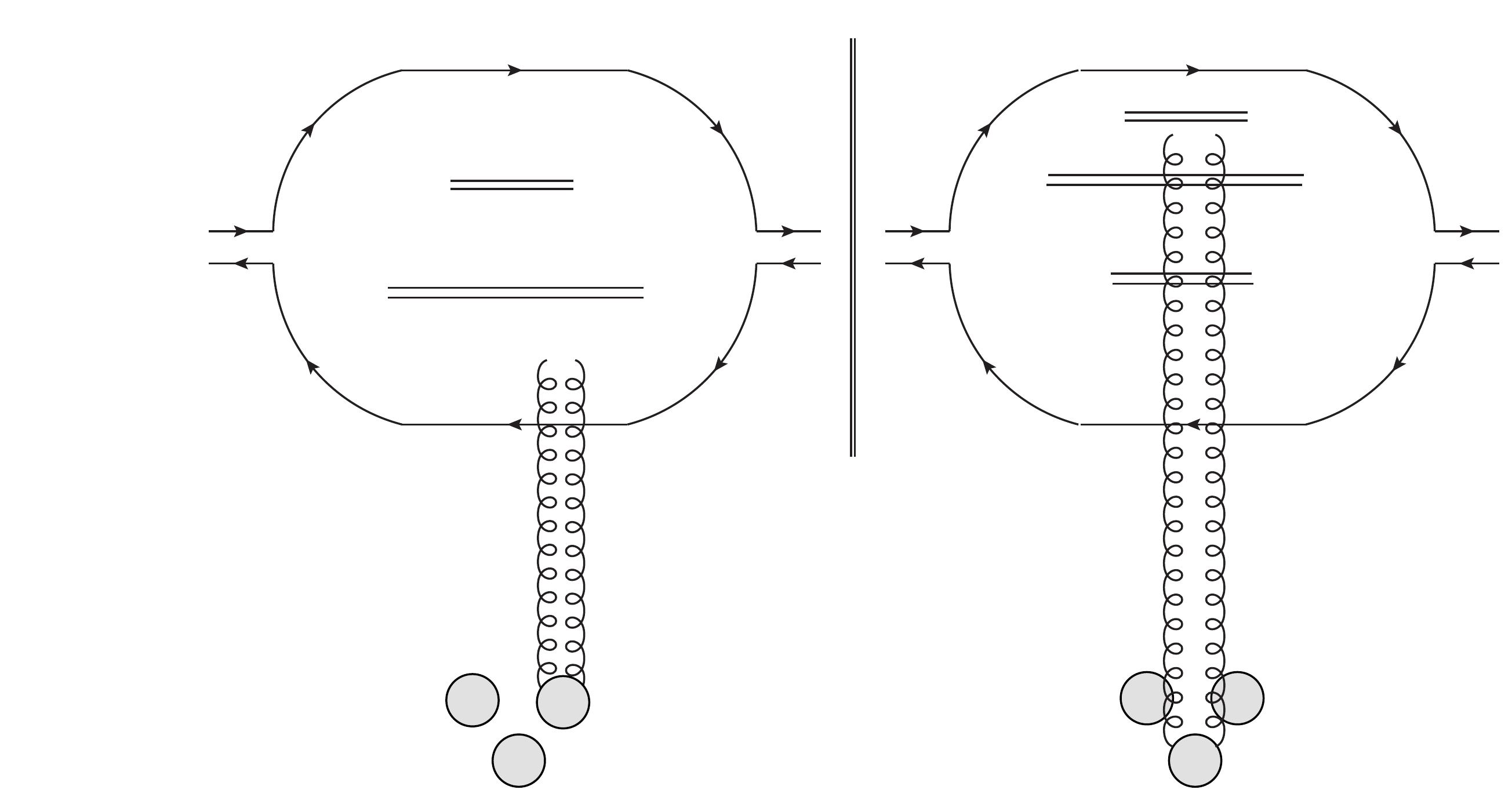}\\
  (a)&(b)
  \end{tabular}
  \end{center}
  \caption{\small  Picture of diagrams that lead to
    (a) onium diffractive dissociative and (b) elastic events.
    The graphs displayed contribute to $\left\langle T^2\right\rangle$ and
    to $\left\langle T\right\rangle^2$ respectively.
    In each graph,
    the double vertical line denotes the final state (up to hadronization),
    the subgraph to the left of it is the amplitude, and the one to the
    right the complex-conjugate amplitude.}
\end{figure}

The cross section for the diffractive dissociation
of a dipole of initial size $x_{01}$ in its scattering off a nucleus
at a fixed impact parameter and with a gap of {\it minimum size $y_0$}
reads in the Good-Walker picture
\be
{\sigma_\text{diff}}=
\sum_X {}_{\tilde y_0}\langle x_{01}|
{\mathbb T}^\dagger(y_0) |X\rangle\langle X| {\mathbb T}(y_0)
|x_{01}\rangle_{\tilde y_0}
-\left|
{}_{\tilde y_0}\langle x_{01}|
{\mathbb T}(y_0)
|x_{01}\rangle_{\tilde y_0}
\right|^2\ ,
\label{eq:goodwalker}
\ee
where $|x_{01}\rangle_{\tilde y_0}$ is the state of the initial
onium evolved to the rapidity $\tilde y_0$ and
${\mathbb T}(y_0)$ is the interaction matrix of the onium Fock states
with a nucleus boosted to the rapidity $y_0$.
$X$ is any dipole final state, and the sum over $X$ also
contains an integration over phase space of the type $\int\prod d^2x_a dz_a$,
where $x_a$ and $z_a$
are respectively the transverse position and the longitudinal momentum fraction
of the gluons (i.e. endpoints of the dipoles).
We refer the reader to~\cite{Munier:2003zb} for an earlier application of this formula
to diffractive deep-inelastic scattering, limited however to a two-dipole,
i.e. $q\bar q g$, diffractive system. The discussion in the more recent paper
of Hatta et al. addresses diffractive systems composed of any number
of gluons~\cite{Hatta:2006hs}, and is very close to our present discussion.

The important point to notice is that in a high-energy scattering, ${\mathbb{T}}(y_0)$
is diagonal in coordinate space. In other words, dipoles are eigenstates
of the interaction. Hence, introducing a complete set of dipoles, using the latter property and
writing the wavefunction of the onium at rapidity $\tilde y_0$ on the state $|X\rangle$ as
$\psi_X(x_{01},\tilde y_0)\equiv\langle X|x_{01}\rangle_{\tilde y_0}$,
Eq.~(\ref{eq:goodwalker}) becomes
\be
{\sigma_\text{diff}}=
\sum_X |\psi_X(x_{01},\tilde y_0)|^2
|\langle X|{\mathbb T}(y_0) |X\rangle|^2
-\left|\sum_X |\psi_X(x_{01},\tilde y_0)|^2
\langle X|{\mathbb T}(y_0)|X\rangle\right|^2.
\ee
We denote by $T(\{r_k\},y_0)\equiv\langle X|{\mathbb T}(y_0) |X\rangle$ the (diagonal)
matrix element of ${\mathbb T}(y_0)$
for a state $|X\rangle$ corresponding to the set $\{r_k\}$ of dipoles
(the $r_k$'s are their sizes, which are enough to characterize them completely for our purpose).
It is also useful to introduce a notation for the averaging
over the realizations of the Fock state of the onium
at rapidity $\tilde y$. For a general function $f(X)$ of the dipole state $|X\rangle$, we write
\be
\sum_X|\psi_X(x_{01},\tilde y)|^2 f(X)=\langle f(X) \rangle_{x_{01},\tilde y}.
\ee
With these notations, the diffractive cross section reads
\be
\sigma_\text{diff}
=\left\langle T^2(\{r_k\},y_0)\right\rangle_{x_{01},Y-y_0}
-\left\langle T(\{r_k\},y_0)\right\rangle_{x_{01},Y-y_0}^2.
\label{eq:sigmadiff}
\ee
We further note that
\be
T(\{r_k\},y_0)=1-\prod_k S(r_k,y_0),
\ee
where $S(r,y)$ is the solution of BK equation~(\ref{eq:BK_S})
for the $S$-matrix element corresponding to the forward elastic
scattering of a dipole of size $r$ at rapidity $y$.

Finally, Eq.~(\ref{eq:sigmadiff})
can trivially be rewritten with the help of $S$ solution of the BK equation
as
\be
\sigma_\text{diff}
=\left\langle \prod_k\left[S(r_k,y_0)\right]^2\right\rangle_{x_{01},Y-y_0}
-\left\langle \prod_k S(r_k,y_0)\right\rangle_{x_{01},Y-y_0}^2.
\label{eq:sigmadiff_S}
\ee

It is useful to note that the second terms in
Eqs.~(\ref{eq:sigmadiff}) and~(\ref{eq:sigmadiff_S}),
namely $\langle T(\{r_k\},y_0)\rangle^2$ and 
$\left\langle \prod_k S(r_k,y_0)\right\rangle^2$,
are actually independant of $y_0$:
This is a simple consequence of
relativistic invariance.
Since we are interested in the distribution of $y_0$,
hence in the derivative of $\sigma_\text{diff}$ with
respect to $y_0$, all $y_0$-independent terms can be dropped.

One could implement this formula in a computer code literally:
$S(r,y)$ can be computed as a function of $r$ for different values of
the rapidity $y$ by integrating the BK equation using standard methods, while
the Fock states of the onium (over which one needs to average)
can be generated using a Monte Carlo implementation of
the dipole model, such as the one developed in Refs.~\cite{Salam:1996nb,toappear}.
However, there is a way to write fully deterministic equations, which
are relatively easier to integrate numerically.


\subsection{Fully deterministic formulation}

We are now in a position to write the diffractive cross section conditioned to the gap $y_0$
as a system of BK equations.

Starting from the previous discussion and introducing the notation
\be
S_2(x_{01},\tilde y)=
\left\langle
\prod_k [S(r_k,y_0)]^2
\right\rangle_{x_{01},\tilde y},
\label{eq:S2}
\ee
we write
\be
\frac{d\sigma_\text{diff}}{dy_0}=-\frac{\partial}{\partial y_0}S_2(x_{01},Y-y_0).
\ee
The crucial observation is that $S_2$ solves the BK equation.
The proof that an expression like
the r.h.s. of Eq.~(\ref{eq:S2}) obeys the BK equation
is very classical in the context
of branching diffusion processes
(see e.g.~\cite{Derrida:1988})
but is less known in the context
of particle physics (see however~\cite{Munier:2014bba} and references therein).
One considers a rapidity interval $\tilde y+d\tilde y$
which one decomposes in an infinitesimal interval $d\tilde y$, over which the initial
onium may either {\it (i)} split to two dipoles with the
probability~(\ref{eq:dipole_splitting}), or {\it (ii)} remain one dipole, with
the complementary probability. The evolution over the remaining rapidity
interval $\tilde y$ produces either
two {\it independent} sets of dipoles $\{r_{k_1}^{(1)}\}$
and $\{r_{k_2}^{(2)}\}$ (case {\it (i)})
or one single set
$\{r_k\}$ (case {\it (ii)}). In equations:
\begin{multline}
S_2(x_{01},\tilde y+d\tilde y)=
\bar\alpha \,d\tilde y\int \frac{d^2 x_2}{2\pi}\frac{x_{01}^2}{x_{02}^2 x_{12}^2}
\left\langle
\prod_{k_1} [S(r_{k_1}^{(1)},y_0)]^2
\right\rangle_{x_{02},\tilde y}
\left\langle
\prod_{k_2} [S(r_{k_2}^{(2)},y_0)]^2
\right\rangle_{x_{12},\tilde y}\\
+\left(
1-\bar\alpha\, d\tilde y
\int \frac{d^2 x_2}{2\pi}\frac{x_{01}^2}{x_{02}^2 x_{12}^2}
\right)\left\langle
\prod_k [S(r_k,y_0)]^2
\right\rangle_{x_{01},\tilde y}.
\end{multline}
Taking the limit $d\tilde y\rightarrow 0$ and performing the
appropriate replacements, we arrive indeed at the
BK equation:
\be
\frac{\partial}{\partial\tilde y}S_2(x_{01},\tilde y)=
\bar\alpha
\int\frac{d^2x_2}{2\pi}\frac{x_{01}^2}{x_{02}^2 x_{12}^2}
\left[S_2(x_{02},\tilde y)S_2(x_{12},\tilde y)-S_2(x_{01},\tilde y)\right].
\ee
The initial condition for $S_2$ reads
\be
S_2(x_{01},\tilde y=0)=[S(x_{01},y_0)]^2,
\ee
where $S$ also solves the BK equation~(\ref{eq:BK_S}), with as an
initial condition the $S$-matrix element for the elastic scattering of a dipole
off a large nucleus at zero rapidity~(\ref{eq:MV_S}).
We recognize the equations first established in Ref.~\cite{Kovchegov:1999ji}
(for a formulation closer to ours,
see Ref.~\cite{Kovchegov:2012mbw}, Sec.~7.2.2, and Ref.~\cite{Hatta:2006hs},
Sec.~2).
Our present work may thus also be understood as a solution to these equations
endowed with an appealing interpretation.

Formulating the diffractive cross section in this way does a priori
not help to address the analytical problem. But
this set of BK equations (for $S$ and for $S_2$)
is straightforward to implement and to
solve numerically. Our most accurate
calculation of the distribution of the rapidity
gap is based on this method.


\section{\label{sec:numerics}Numerical calculations}

\subsection{Diffraction as the solution of a deterministic equation}

We can obtain $d\sigma_\text{diff}/dy_0$ for a given $y_0$
from the Good-Walker formula, integrating numerically the BK equations
that follow.
Such calculations were done in Ref.~\cite{Levin:2001yv,Levin:2001pr,Levin:2002fj},
but with a quite different scope.

We use a code that solves the BK equation at fixed impact parameter and
in momentum space which is very similar to BKSolver
developed in Ref.~\cite{Enberg:2005cb}.
We shall explain very concretely how we operated it to compute
$d\sigma_\text{diff}/dy_0$. 

We start with a McLerran-Venugopalan
initial condition for the elastic amplitude,
\be
T_\text{MV}(r)=1-S_\text{MV}(r),
\ee
where $S_\text{MV}$ was given in Eq.~(\ref{eq:MV_S}),
and convert it to momentum space. Generically, the transformation that
realizes the mapping reads
\be
\tilde T(k,y)=\int \frac{d^2 r}{2\pi r^2}e^{ik\cdot r}T(r,y).
\ee
We then evolve $\tilde T_\text{MV}$ numerically to the rapidity~$y_0$.
Then, we Fourier transform $\tilde T(k,y_0)$ back to coordinate space,
compute the quantity $2T(r,y_0)-T^2(r,y_0)$ and
evolve its Fourier transform for $\tilde y_0=Y-y_0$ more units of rapidity.
We repeat this procedure for different values of $y_0$ ranging between
$0$ and $Y$.
The result is then expressed in coordinate space,
and its $y_0$-derivative (up to a sign) is plotted
as a function of $y_0$ for selected values of the initial dipole size $x_{01}$.

We consider two different values of the total rapidity:
A phenomenologically realistic one,
$\bar\alpha Y=3$, and a larger value, $\bar\alpha Y=20$,
which enables us to better approach the asymptotics in order
to test more quantitatively our picture, which should be formally exact at 
asymptotic rapidities.
We chose to set $\Lambda=200\ \text{MeV}$ and $Q_\text{MV}=1\ \text{GeV}$.
In what follows, the momenta and distances will always be expressed in
units of $Q_\text{MV}$ and of $Q^{-1}_\text{MV}$ respectively.

We evolve the scattering amplitude of a dipole of size $x_{01}$
with the nucleus through the BK equation. We first
measure the saturation scale $Q_s(Y)$ at rapidity $Y$
defining it as $S(x_{01}=2/Q_s(Y),Y)=\frac12$.
For $\bar\alpha Y=3$, we find $Q_s(\bar\alpha Y=3)\simeq 230$
and for $\bar\alpha Y=20$,
$Q_s(\bar\alpha Y=20)\simeq 2\times 10^{19}$, consistently in order of
magnitude with the asymptotic
formula
$Q_s^2(Y)\simeq Q_\text{MV}^2e^{\bar\alpha Y\chi'(\gamma_0)}/(\bar\alpha Y)^{3/2}$
which gives $Q_s(\bar\alpha Y=3)\simeq 665$ and
$Q_s(\bar\alpha Y=20)\simeq 1.7\times 10^{20}$ respectively.\footnote{
  Note that even for $\bar\alpha Y=3$, the values of $Q_s$ we find
  are way too large to be realistic for phenomenology.
  This is because, as well-known,
  the BFKL evolution at leading order predicts a growth of the
  gluon density with the rapidity which is too fast.
  It would be tamed thanks to next-to-leading order corrections.  
  The latter would change the numbers but would not alter our
  picture.}

Then, we compute $\sigma_\text{diff}$ for the two chosen values
of the total rapidity~$Y$, scanning in each case
the interval $y_0\in]0,Y[$.
    We choose $x_{01}$ in the scaling region, and at its border.
We fit the following formula to the numerical result
\be
\frac{d\sigma_\text{diff}}{dy_0}=\text{const}\times
\left(\frac{
  \bar\alpha Y}{\bar\alpha y_0 \bar\alpha(Y-y_0)}
\right)^{3/2}\exp\left\{
-\frac{\ln^2[x_{01}^2 Q_s^2(Y)]}
{2\bar\alpha(Y-y_0)\chi''(\gamma_0)}\right\},
\label{eq:fitted_formula}
\ee
where the overall constant is the only free parameter.

\begin{figure}[h]
  \begin{center}
  \begin{tabular}{cc}
    \includegraphics[width=0.5\textwidth]{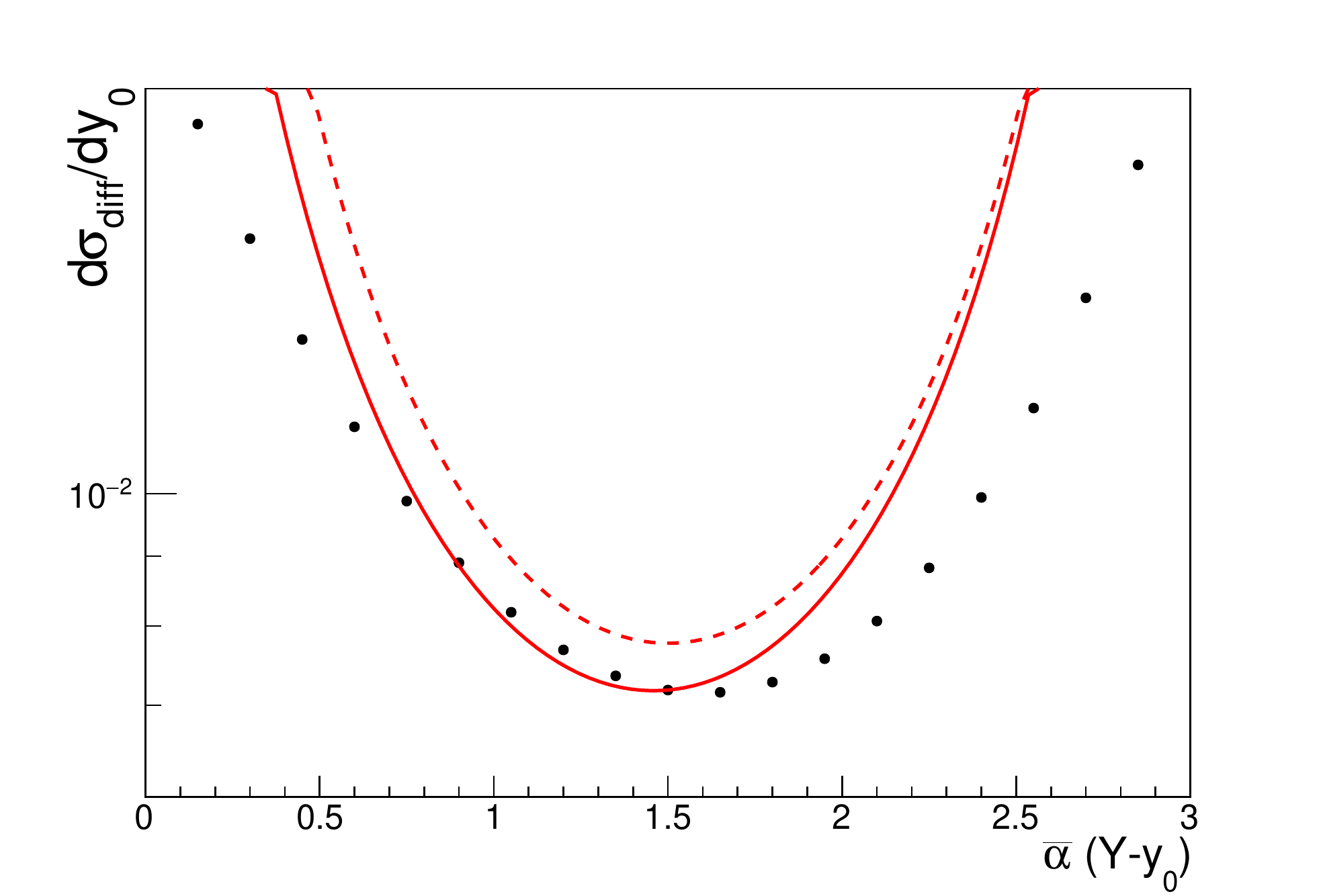}&
    \includegraphics[width=0.5\textwidth]{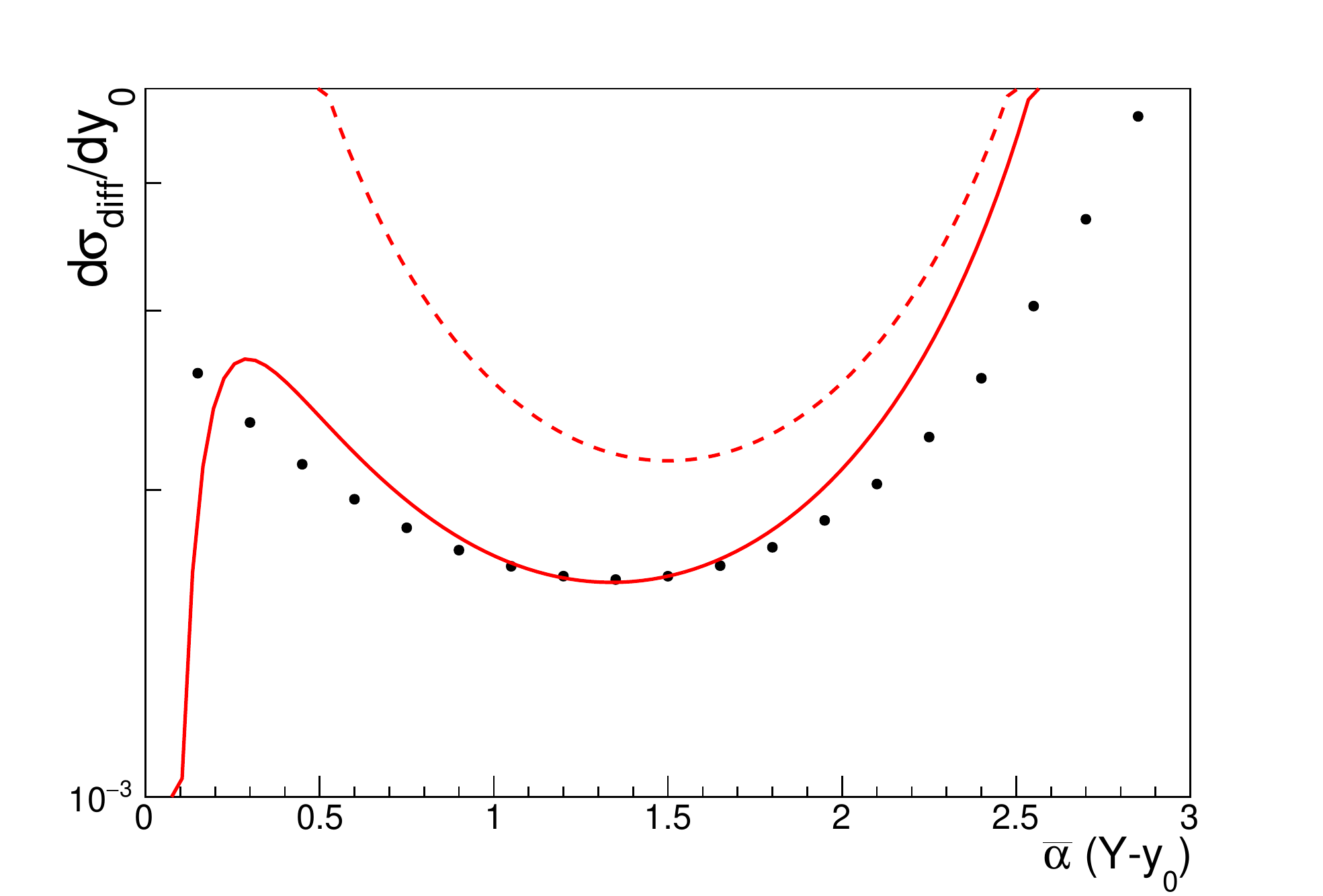}\\
    (a)&(b)
  \end{tabular}
  \end{center}
  \caption{\small \label{fig:distrib_scaling_3}
    Diffractive cross section for $\bar\alpha Y=3$
    as a function of $\bar\alpha(Y-y_0)$ for two
    sizes of the onium chosen in the
    scaling region: (a) $x_{01}=8\times 10^{-4}$,
    (b) $x_{01}=2\times 10^{-4}$
    in units of the inverse McLerran-Venugopalan
    saturation scale $1/Q_\text{MV}$.
    The points correspond to the values of $Y-y_0$ at which
    the diffractive cross section is computed numerically.
    The continuous line is the fit of the theoretical
    formula~(\ref{eq:fitted_formula}). The dashed line is a graph of the
    same formula, without the Gaussian suppression factor,
    and without refitting the parameters. It represents
    the large-$Y$ asymptotics deeply in the scaling region.
    }
\end{figure}

\begin{figure}[h]
  \begin{center}
  \begin{tabular}{cc}
    \includegraphics[width=0.5\textwidth]{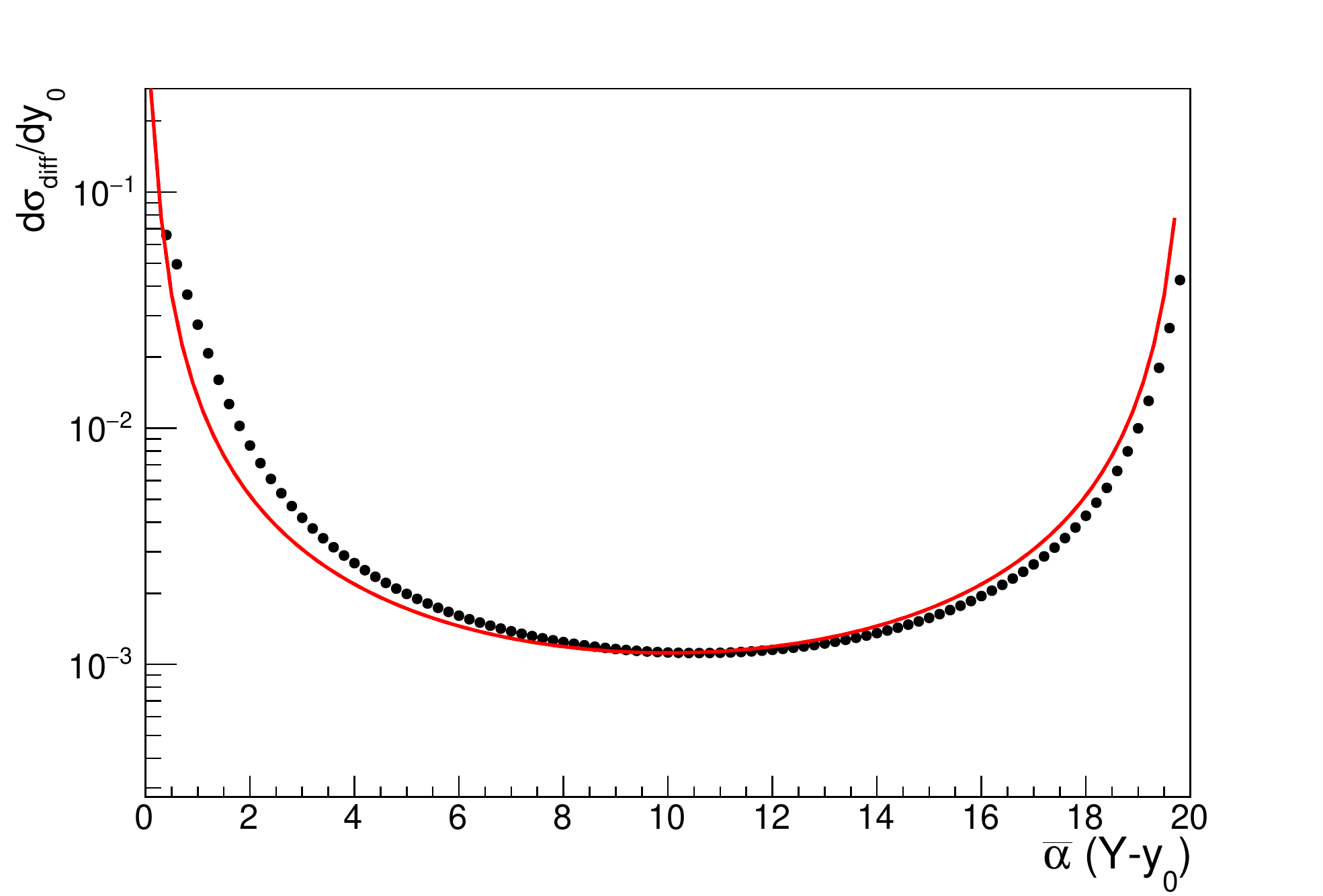}&
    \includegraphics[width=0.5\textwidth]{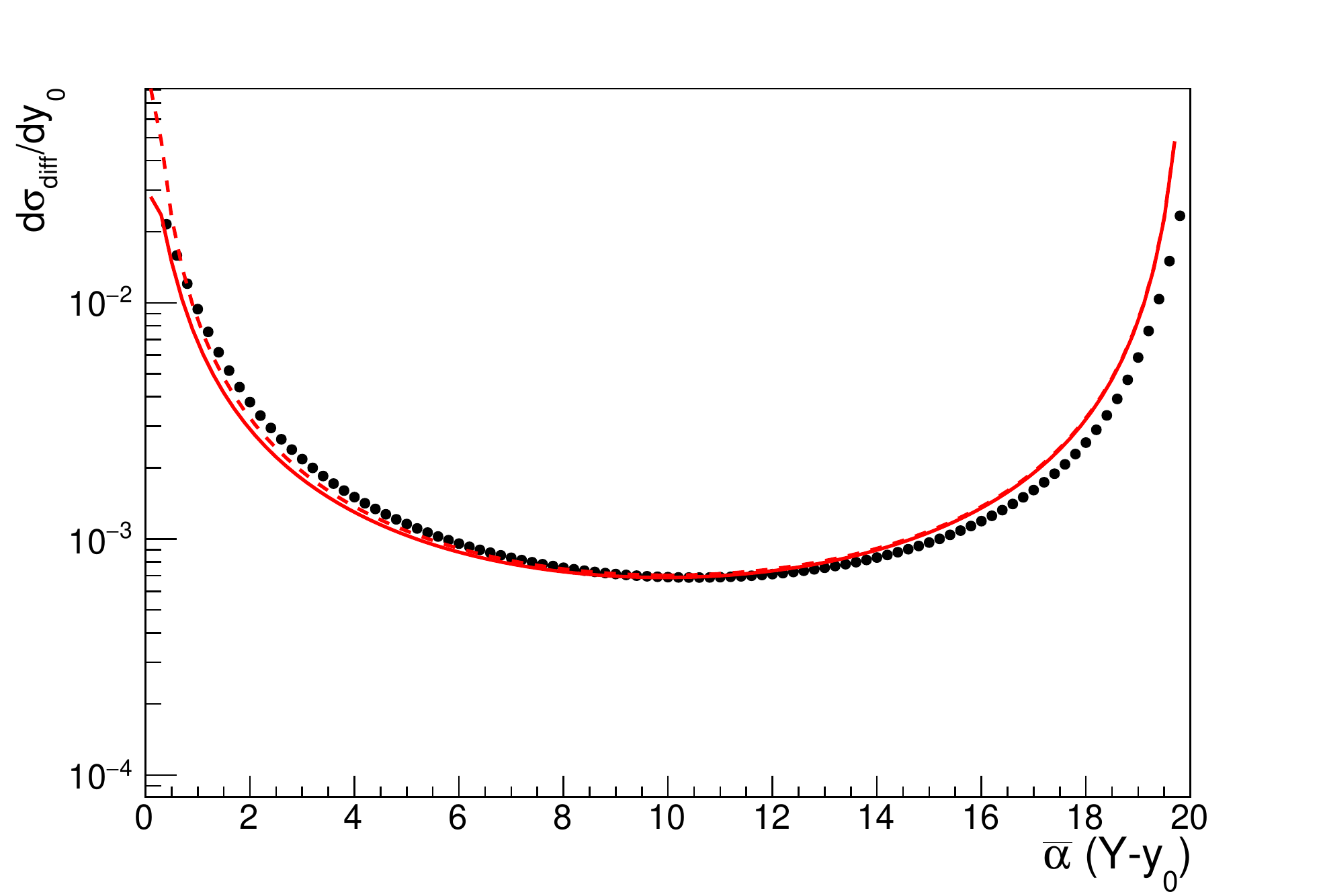}\\
    (a)&(b)\\
        \includegraphics[width=0.5\textwidth]{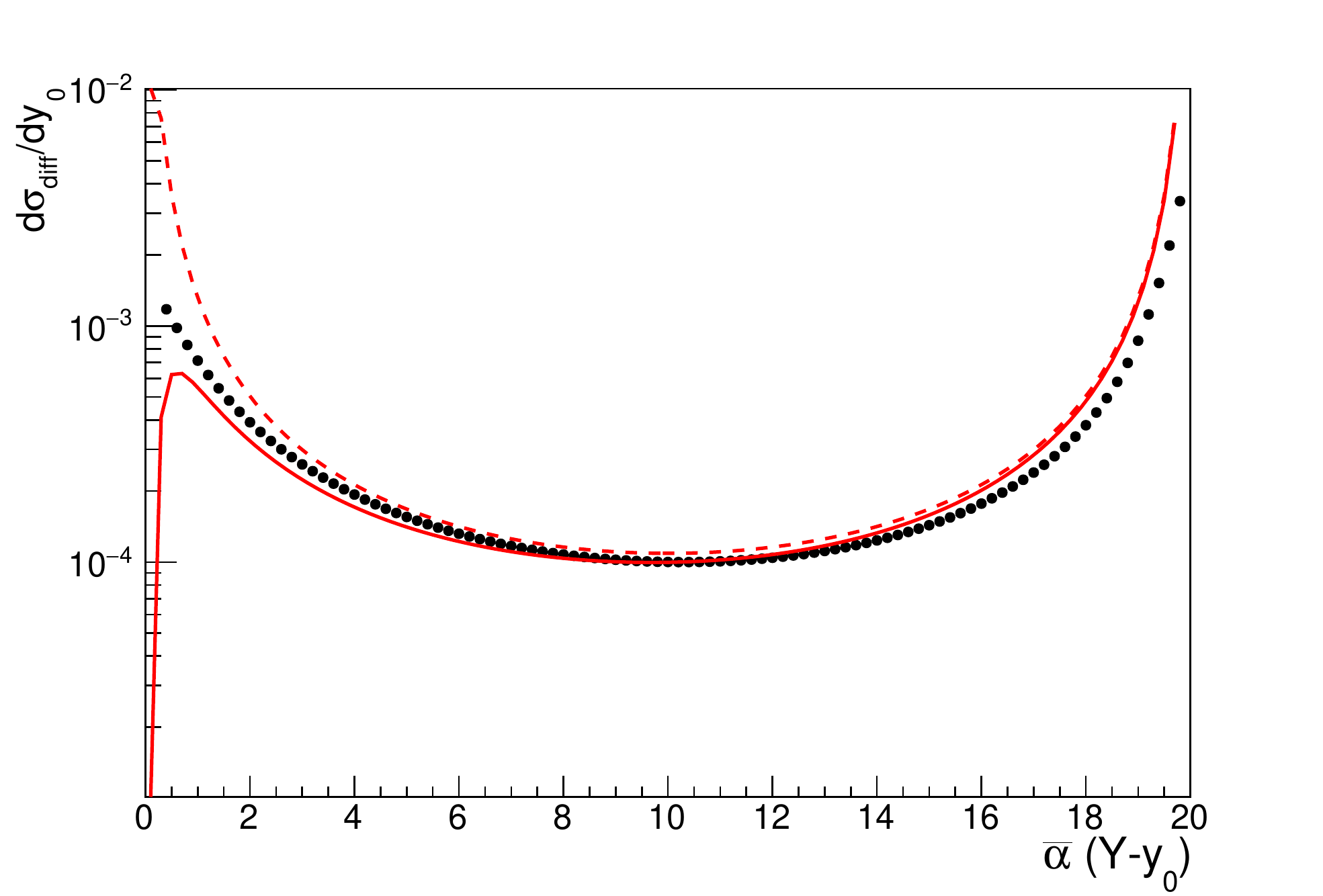}&
    \includegraphics[width=0.5\textwidth]{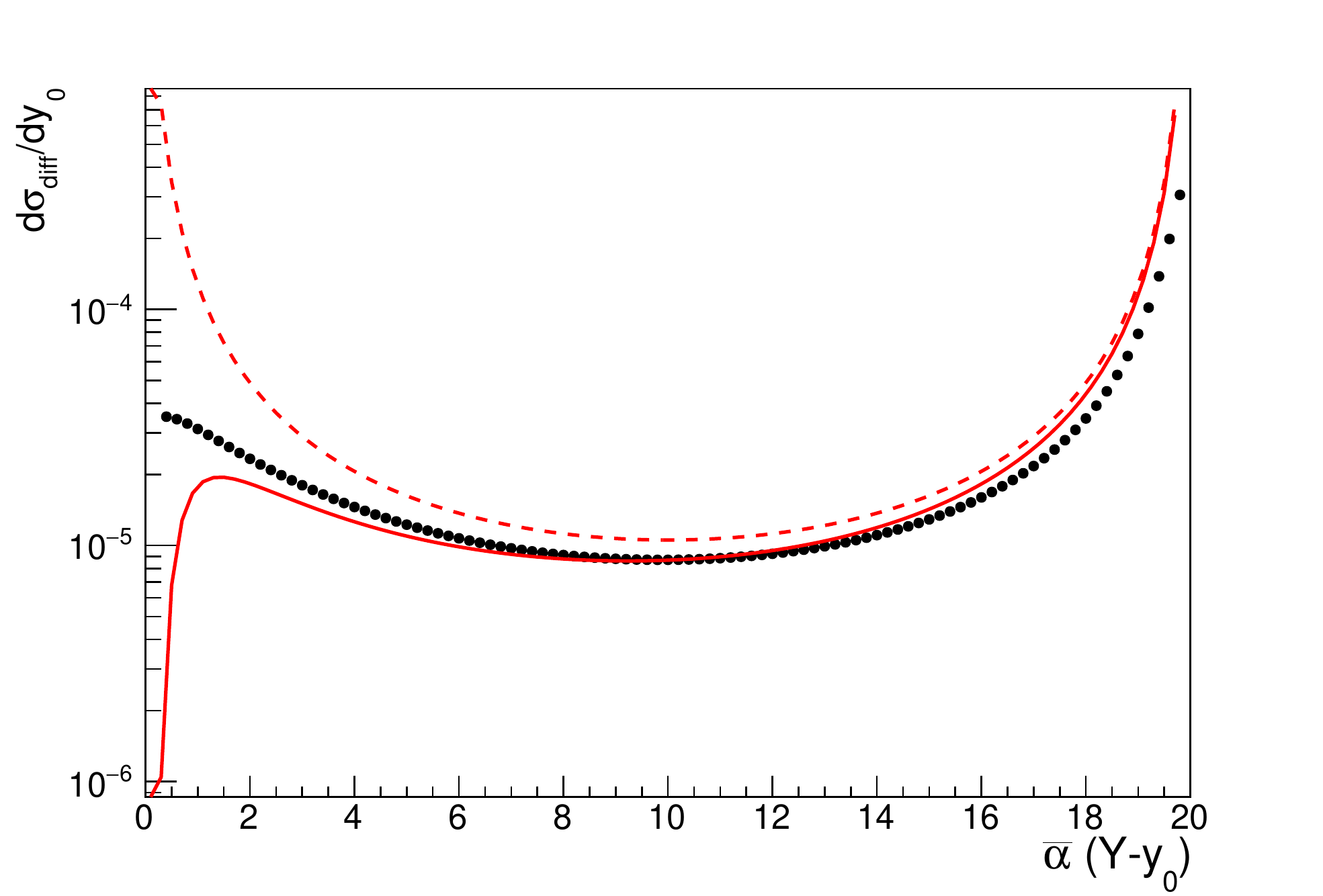}\\
    (c)&(d)
  \end{tabular}
  \end{center}
  \caption{\small \label{fig:distrib_scaling_20}
    The same as in Fig.~\ref{fig:distrib_scaling_3} for $\bar\alpha Y=20$
    and (a) $x_{01}=10^{-19}$, (b) $x_{01}=10^{-20}$,
    (c) $x_{01}=10^{-21}$, (d) $x_{01}=10^{-22}$.
    }
\end{figure}

The results
are shown in Figs.~\ref{fig:distrib_scaling_3},\ref{fig:distrib_scaling_20}.
We see that our asymptotic formula describes
qualitatively the numerical data for $\bar\alpha Y=3$. For
such a low value of the rapidity, the scaling region is very narrow.
Furthermore, the conditions $\bar\alpha y_0>1$ and $\bar\alpha(Y- y_0)>1$
can obviously not be satisfied with such low values of the
rapidity. It is however interesting that although the corrections to
the asymptotics are expected
to be very large for this choice of the rapidity, the general trend
is in agreement with the expectations from the asymptotic formula.

For $\bar\alpha Y=20$ (Fig.~\ref{fig:distrib_scaling_20}),
the numerical data is
remarkably well described in the scaling region and on its border.
The power law~(\ref{eq:main_result}) is beautifully seen
in the strict scaling region,
and the transition to the non-scaling regime is also well described
by our theoretical formula~(\ref{eq:fitted_formula}).
One has to keep in mind
that it was established under the explicit
assumption $\bar\alpha Y\gg 1$,
and thus is expected to
match the regime of asymptotic
rapidity only. Note also
that again, it is not supposed
to be accurate for $\bar\alpha y_0\sim 1$ or $\bar\alpha (Y-y_0)\sim 1$.

Let us comment on the comparison on our results to the numerical calculation
in Ref.~\cite{Levin:2001pr}.
In there, $\sigma_\text{diff}$ was computed at fixed $Y$
(set to~10) as a function of $y_0$. The dependence upon $y_0$
was found close to linear (see Fig.~2 in Ref.~\cite{Levin:2001pr}),
which results  in an almost flat
dependence of $d\sigma_\text{diff}/dy_0$ upon $y_0$.
Since $\bar\alpha Y\simeq 2.8$ in that calculation, this is not
inconsistent with our results: Indeed, for low rapidities,
$d\sigma_\text{diff}/dy_0$ becomes indeed flatter: Subasymptotic terms
screen the asymptotic behavior we have derived here.

\subsection{Implementation of a stochastic formulation of diffraction\label{sec:stochastic_formulation}}

\begin{figure}[h]
  \begin{center}
  \begin{tabular}{cc}
    \includegraphics[width=0.5\textwidth]{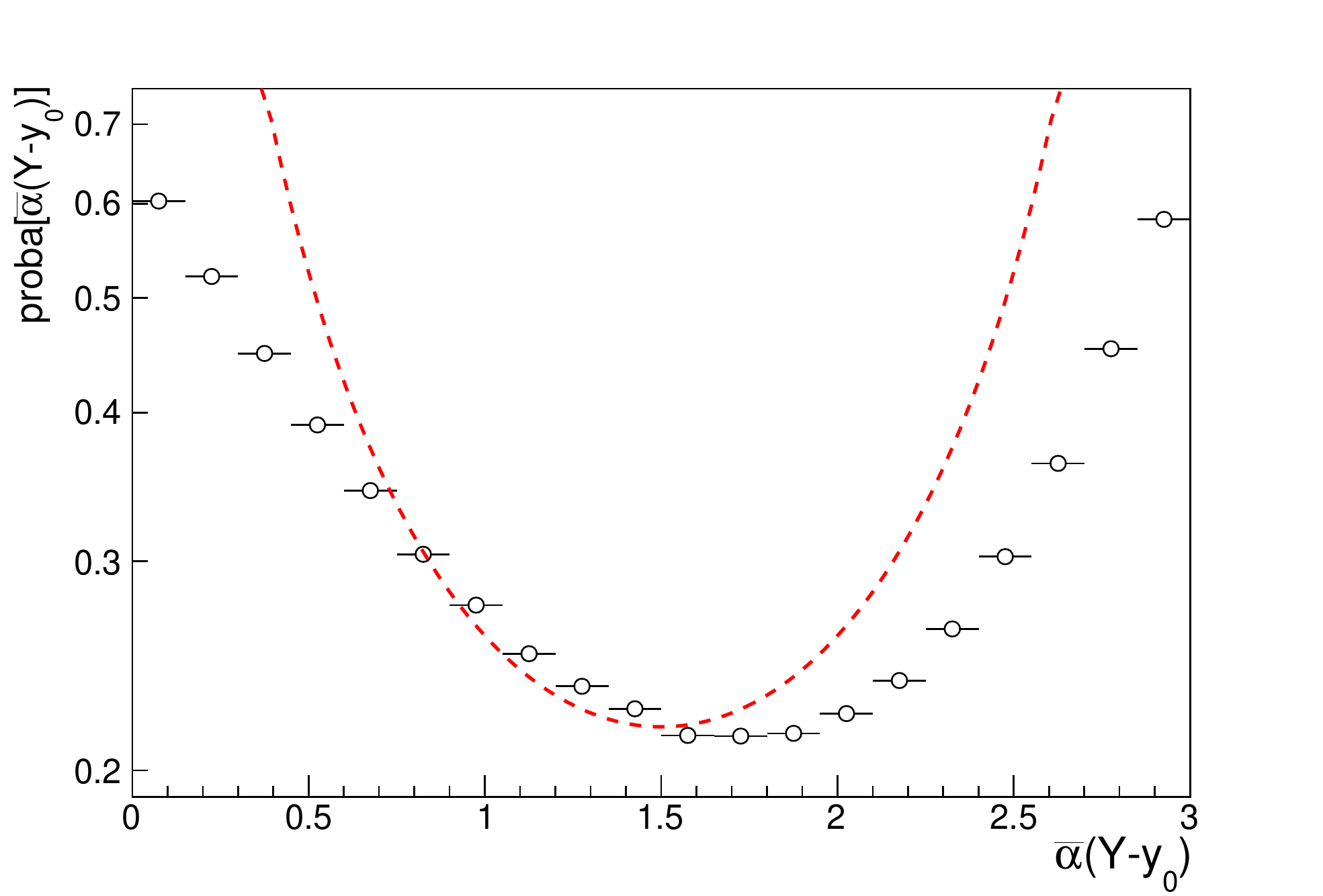}&
    \includegraphics[width=0.5\textwidth]{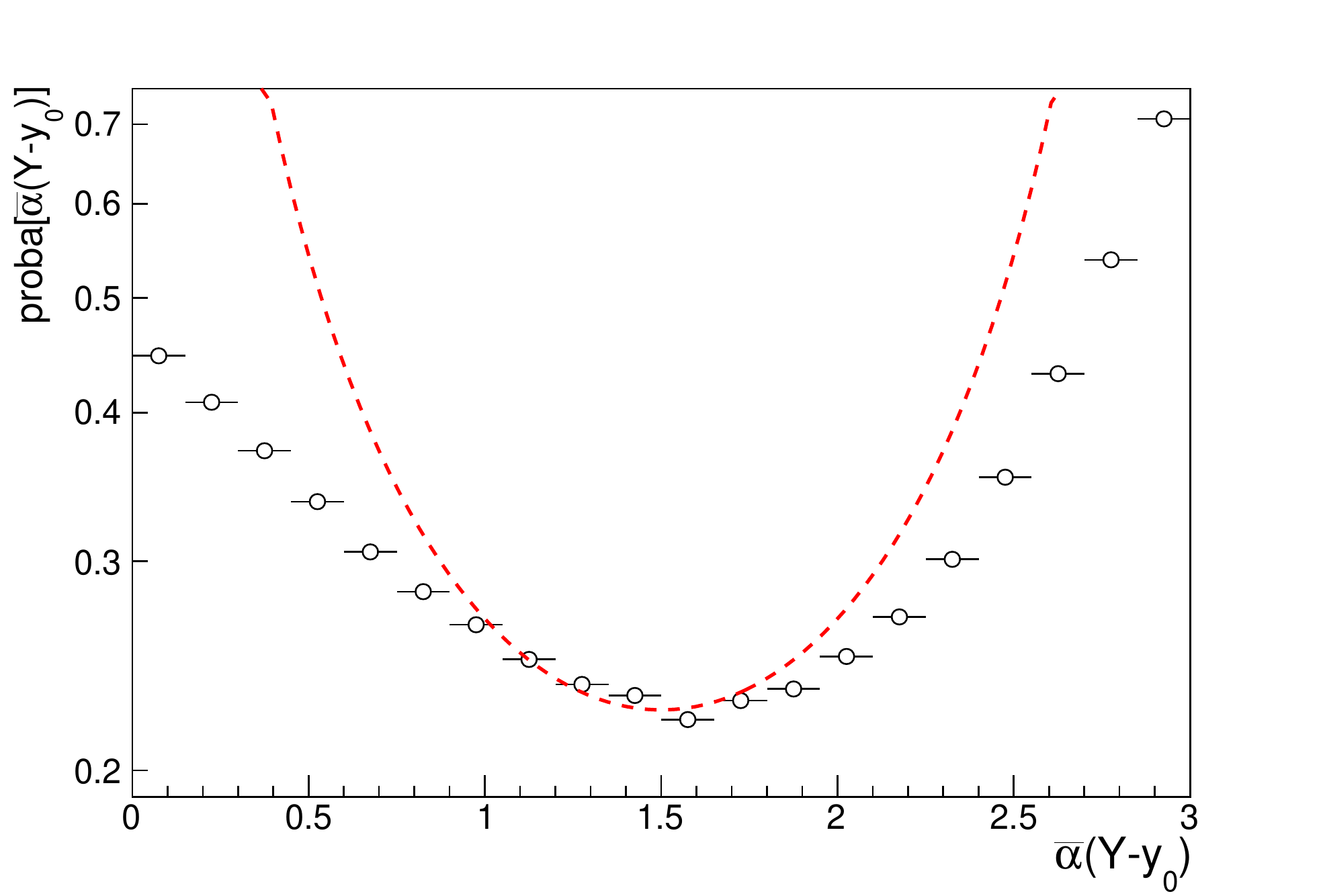}\\
    (a)&(b)
  \end{tabular}
  \end{center}
  \caption{\small \label{fig:stocha}
    Distribution of the rapidity at which the common ancestor
    split for
    $\bar\alpha Y=3$
    and in the scaling region.
    (a) $x_{01}=5\times 10^{-3}$, (b)
    $x_{01}=2.5\times 10^{-3}$.
    The points represent the Monte Carlo calculation as described
    in the text.
    The asymptotic prediction~(\ref{eq:main_result}) is shown (dashed line).
    The overall normalization is chosen to coincide with the numerical
    data at $\bar\alpha(Y-y_0)=1.5$.    
    }
\end{figure}

\subsubsection{Procedure}

In order to test in a more detailed way
our picture of diffraction and its analogy with ancestry problems,
we need to set up a definition for a diffractive event in a Monte Carlo
approach.
This is not straightforward at all because as mentioned
in the introductive parts,
diffraction is a quantum mechanical phenomenon which may a priori not
be defined in purely probabilistic terms.

However, the following procedure is very close to the spirit of the picture
we propose.
Start with one dipole, evolve it to rapidity $y$ using a Monte Carlo
implementation of the dipole model~\cite{Salam:1996nb,toappear}.
One obtains a configuration
of dipoles, which one replicates once: One copy
will build the amplitude, the other one its complex conjugate.
We require that these dipoles scatter elastically.

More precisely, each dipole scatters off the nucleus at rest
with probability $T_\text{MV}$ (or does
not scatter with the complementary probability
$S_\text{MV}=1-T_\text{MV}$). We declare
that an event is diffractive (or purely elastic) if overall
there is at least one dipole which scatters in the amplitude, and
at least one
dipole in the complex-conjugate amplitude.
Now the rapidity gap in that particular event is defined to be
the rapidity (counted from the nucleus) at which the
common ancestor of {\it all} dipoles that scatter split.

A comment is in order. $S$ and $T$ are not probabilities, but probability
amplitudes. In our procedure, we trade them for probabilities,
which is not rigorously
correct. But counting the fraction of events 
which have at least one interaction in the
amplitude and one in the complex-conjugate
amplitude is indeed fully equivalent to computing the diffractive
cross section using the Good-Walker formula.

\subsubsection{Results}

Many events are needed in order to arrive at
a good statistical accuracy on the rapidity of the common ancestors
(about $10^7$ realizations of the dipole evolution are in order).
In order to lower the complexity and ease the calculation,
we simplify the procedure outlined above by replacing
the McLerran-Venugopalan amplitude by a sharp $\theta$ function:
The dipoles which have a size larger than $1/Q_\text{MV}$
interact, the ones that have a size smaller do not, with unit probability
in each case.
This should be a good approximation since as well-known, the MV amplitude
is much steeper than the distribution of dipole sizes generated by the BFKL
evolution of the onium.
In any  implementation of the dipole model, an (unphysical) ultraviolet cutoff
is required to regularize the collinear singularity.
We choose
it to be 10\% of the size of the initial onium (which is quite large)
in order to keep the
number of spectator dipoles reasonable. This has the effect
of decreasing the growth of the saturation scale with the rapidity.
But we believe that
this should not change qualitatively the distribution of common
ancestors, since we eventually probe the onium Fock state in size regions
which are much larger than this ultraviolet cutoff.

The results are shown in Fig.~\ref{fig:stocha} for $\bar\alpha Y=3$.
We see that the distribution of the splitting rapidity of the most recent
common ancestor follows indeed the same trend as the rapidity gap distribution
in Fig.~\ref{fig:distrib_scaling_3},
consistently with our picture of diffraction.


\section{Conclusion}

The main practical result of this paper is a robust
prediction for the distribution
of the rapidity gaps in diffractive deep-inelastic
scattering of a large nucleus
at very high energy, see Eq.~(\ref{eq:main_result}) for the simplest
expression,
or~(\ref{eq:diff}) for an expression
which has a wider range of applicability.
While the strict asymptotic form we have obtained is
probably out of experimental reach, our numerical simulations suggest that
the global shape of this distribution
may show up already at realistic values of the rapidity,
maybe attainable at a
future electron-ion collider.

Our main thrust was actually the theoretical understanding of the
microscopic mechanism behind high-mass diffractive dissociation
phenomena. We have found that the diffractive events were generated
by unusually large fluctuations in the partonic evolution of
the initial onium.
The rapidity $y_0$
(when counted from the nucleus)
at which they occur determines the size of the gap.

Interestingly enough, the distribution of the rapidity gap
size $y_0$ has exactly the same form as
the distribution of the decay time of the common ancestor
of a set of extreme
particles in a branching diffusion process.
It is actually the mechanism how
the common ancestor is singled out in branching random walks
which directly motivated our calculation in the context of
particle physics.
This exciting analogy between a purely probabilistic problem (ancestry
in branching random walks) and
a process in which quantum mechanics is essential (diffraction)
may probably be pushed further.
Indeed, the calculation we have performed
lead to the same result as Derrida and Mottishaw
but is not quite formulated in the same way.
The analogy
between the ancestry problem and diffraction would deserve
more studies.

Among other open problems,
taking into account formally subleading effects in the
dipole evolution such as the
running of the QCD coupling would be in order
to be able to appreciate quantitatively how relevant our
result is for the phenomenology at a electron-ion collider.


\section*{Acknowledgements}

The  work  of  AHM  is  supported in part by
the U.S. Department of Energy Grant
\# DE-FG02-92ER40699.
The work of SM is supported in part by the Agence Nationale
de la Recherche under the project \# ANR-16-CE31-0019.


\appendix

\section{Proof that a diffractive event
  is triggered by a tip fluctuation}

The goal of this appendix is to exhibit an argument that when there is (at least) a dipole
larger than $1/Q_s(y_0)$ in the onium at rapidity $\tilde y_0$, it stems predominantly
from a front fluctuation (happening, by definition,
in the beginning of the evolution)
followed by a tip fluctuation occurring precisely at $\tilde y_0$.
We are going to show that earlier fluctuations, at $\tilde y_1<\tilde y_0$ which contain
dipoles larger than $1/Q_s(y_1)$ cannot generate dipoles of size greater than $1/Q_s(y_0)$
at rapidity $\tilde y_0$.

In order for a rare fluctuation occurring at
rapidity $\tilde y_1=Y-y_1$ (with respect to the onium)
to generate offspring
in the saturation region at $\tilde y_0>\tilde y_1$,
it needs to consist in one or a few dipoles of size larger
than $1/Q_s(y_1)$ by at least some factor
$e^{\delta/2}$, where $\delta$
is a positive number that we shall evaluate soon.

The probability that there be at least one dipole
larger than $e^{\delta/2}/Q_s(y_1)$ in the set of dipoles present
in the Fock state at rapidity
$\tilde y_1$ solves the BK equation, and thus reads, in the scaling region
\begin{multline}
P(x_{01},\tilde y_1|e^{\delta/2}/Q_s(y_1))=c_P \ln
\frac{e^\delta}{x_\perp^2(\tilde y_1)Q_s^2(y_1)}
\left[{x_\perp^2(\tilde y_1)Q_s^2(y_1)}
  \right]^{\gamma_0}\\
\times
\exp\left\{
  -\frac{\ln^2 [x_\perp^2(\tilde y_1)Q_s^2(y_1)]}
  {2\bar\alpha\tilde y_1\chi''(\gamma_0)}
  \right\}
\end{multline}
(see Eq.~(\ref{eq:Pscaling})) which may be approximated by
\be
c_P \ln \frac{1}{x_{01}^2Q_s^2(Y)}
\left[{x_{01}Q_s^2(Y)}
  \right]^{\gamma_0}
e^{-\gamma_0\delta}
\left[
  \frac{\bar\alpha Y}{\bar\alpha y_1\bar\alpha\tilde y_1}
  \right]^{3/2}
\exp\left\{
  -\frac{\ln^2 [x_{01}^2Q_s^2(Y)]}
  {2\bar\alpha \tilde y_1\chi''(\gamma_0)}
  \right\},
  \label{eq:P_appendix}
\ee
where we have used the same inequality as in Eq.~(\ref{eq:approx_log})
with the replacement $y_0\rightarrow y_1$, and we have anticipated
that the relevant values of $\delta$ will be small compared to
$\left|\ln[x_{01}^2 Q_s^2(Y)]\right|$.

Let us focus on an event which has its tip at $e^{\delta/2}/Q_s(y_1)$
at the rapidity $\tilde y_1$.
Then at rapidity $\tilde y_0$, the dipoles at the tip of the front
generated by the latter
have squared size
given typically by the mean position of the tip
of a front starting with one dipole
at rapidity $\tilde y_1$ and evolved to rapidity $\tilde y_0$.
It is given by $x_\perp^2$ in Eq.~(\ref{eq:x_perp}), with the substitutions
\be
\tilde y\rightarrow \tilde y_0-\tilde y_1=y_1-y_0
\quad\text{and}\quad
x_{01}^2\rightarrow \frac{e^\delta}{Q_s^2(y_1)}.
\ee
The squared size of the largest dipole at rapidity $\tilde y_0$
thus reads
\be
\frac{e^\delta}{Q_s^2(y_1)}
\frac{e^{\bar\alpha(y_1-y_0)\chi'(\gamma_0)}}
     {\left[\bar\alpha(y_1-y_0)\right]^{3/2\gamma_0}}.
\ee
We require that this size be at least as large as $1/Q_s^2(y_0)$,
which is the condition for having a gap covering a rapidity
interval larger than $y_0$ in the event.
Straightforward algebra leads to the condition
$\delta\geq\delta_0$, where
\be
\delta_0\equiv\frac{3}{2\gamma_0}
\ln\frac{\bar\alpha(y_1-y_0)\bar\alpha y_0}{\bar\alpha y_1}.
\ee
Hence the probability that there be a fluctuation beyond the nuclear
saturation boundary at rapidity
$\tilde y_1$ which results in dipoles in the saturation
region at $\tilde y_0$ is $P$ in Eq.~(\ref{eq:P_appendix}) with
$\delta\rightarrow\delta_0$. The
ratio of the latter to the probability 
of finding dipoles in the saturation region at $y_0$ reads
\be
\frac{P(x_{01},\tilde y_1|e^{\delta_0/2}/Q_s(y_1))}{P(x_{01},\tilde y_0|1/Q_s(y_0))}
=\left[\frac{\bar\alpha(Y-y_0)}{\bar\alpha(y_1-y_0)\bar\alpha(Y-y_1)}\right]^{3/2}
\exp\left\{
-\frac{\ln^2[x_{01}^2 Q_s^2(Y)]}{2\bar\alpha(Y-y_1)\chi''(\gamma_0)}
\right\},
\ee
when we remain in the scaling region.
This expression contains all $y_1$-dependent
factors of ${P(x_{01},\tilde y_1|e^{\delta_0/2}/Q_s(y_1))}$.

As the fluctuation may happen at any rapidity between say $y_0+\Delta$
and $Y$, a conservative
upper bound on the fraction of realizations which contain dipoles
larger than $1/Q_s(y_0)$ at $y_0$ and dipoles larger than
$1/Q_s(y_1)$ at $y_1$ with respect
to the total number of realizations which have dipoles
larger than $1/Q_s(y_0)$ at $y_0$ reads
\be
R=\int_{\bar\alpha(y_0+\Delta)}^{\bar\alpha Y} d(\bar\alpha y_1)
\frac{P(x_{01},\tilde y_1|e^{\delta_0/2}/Q_s(y_1))}{P(x_{01},\tilde y_0|1/Q_s(y_0))}.
\ee
We can scale out the factor $[\bar\alpha(Y-y_0)]^{-1/2}$, which is much less than~1
since $y_0$ cannot be close to $Y$ by assumption.
The change of variable $x=(y_1-y_0)/(Y-y_0)$ brings the integral into the form
\be
R=\frac{1}{\sqrt{\bar\alpha(Y-y_0)}}
\int_{\Delta/(Y-y_0)}^1\frac{dx}{x^{3/2}(1-x)^{3/2}}
\exp\left\{
-\frac{\ln^2[x^2_{01} Q_s^2(Y)]}{2\chi''(\gamma_0)\bar\alpha(Y-y_0)(1-x)}
\right\}.
\ee
The integrand of the remaining integral has two singularities which
may potentially enhance the integral, located at the edges of the
integration region.
The singularity at $y_1\rightarrow Y$ results in the following contribution
to the integral over $x$ appearing in~$R$:
\be
\int^1 \frac{dx}{(1-x)^{3/2}}\exp\left\{
-\frac{\ln^2[x^2_{01} Q_s^2(Y)]}{2\chi''(\gamma_0)\bar\alpha(Y-y_0)(1-x)}
\right\}
\ee
The lower bound can be set to $-\infty$, and the integral may then be computed
exactly. Inserting the result into the expression of $R$, we get the
following contribution:
\be
R\xrightarrow[{|\ln[x_{01}^2Q_s^2(Y)]|\ll\sqrt{\bar\alpha \tilde y_0\chi''(\gamma_0)}}]{}
\frac{\sqrt{2\pi\chi''(\gamma_0)}}{\left|\ln [x_{01}^2 Q_s^2(Y)]\right|}
\ee
which is parametrically small compared to unity by assumption.
The contribution of the singularity at $y_1\rightarrow y_0$ (namely $x\rightarrow 0$)
is evaluated in a similar way:
\be
R\xrightarrow[\Delta\ll\tilde y_0]{}
\frac{1}{2\sqrt{\bar\alpha\Delta}}.
\ee
This is of order unity only for $\bar\alpha\Delta\sim 1$.
This proves that the fluctuations that contribute to the production of
dipoles larger than $1/Q_s(y_0)$ at $\tilde y_0$
happen indeed close to the rapidity $\tilde y_0$: This is a tip fluctuation.

The reason why this check is crucial to justify our picture and
its quantitative form Eq.~(\ref{eq:sigma_diff=P}) is the following.
According to our discussion in Sec.~\ref{sec:el_diff}, if a dipole is produced
in the saturation region of the nucleus at rapidity~$\tilde y_1$, then the gap
has a size which a priori is at least $y_1$. If in the same event there were
dipoles in the nuclear saturation region at rapidity~$\tilde y_0$, then this
event would count as a diffractive event of gap size {\it exactly} equal to~$y_0$
according to Eq.~(\ref{eq:sigma_diff=P}).
This would be contradictory, unless $\tilde y_1\simeq \tilde y_0$, which is what
we have just proven.


\end{document}